\documentclass[a4,notitlepage,12pt]{jedm}

\usepackage{graphicx}
\usepackage{amssymb}
\usepackage{hyperref}

\usepackage{lineno}
\newcommand{\name}{RiPPLE\xspace}
\usepackage[linesnumbered,ruled]{algorithm2e}
\usepackage{xspace}
\usepackage{caption}
\usepackage{subcaption}

\begin{document}


\title{Recommendation in Personalised Peer-Learning Environments}



\author{{\large Hassan Khosravi}\\University of Queensland\\h.khosravi@uq.edu.au}
\date{}
\maketitle

\begin{abstract}
Recommendation in Personalised Peer Learning Environments (\name) is an adaptive, crowdsourced, web-based, student-facing, open-source platform that employs exemplary techniques from the fields of machine learning, crowd-sourcing, learning analytics and recommender systems to provide personalised content and learning support at scale. \name presents students with a repository of tagged multiple-choice questions and provides instant feedback in response to student answers. The repository of the questions is created in partnership with the students through the use of crowdsourcing. \name  uses students’ responses to the questions to approximate their knowledge states. Based on their knowledge state and learning needs, each student is recommended a set of personalised questions. For students that are interested in providing learning support, seeking learning support or finding study partners, \name recommends peer learning sessions based on their availability, knowledge state and preferences.  
This paper describes the \name  interface and an implementation of that interface that has been built at the University of Queensland. The \name platform and a reference implementation are released as an open-source package under the Apache 2.0 license via GitHub.
\end{abstract}
%
%
\newpage
\tableofcontents
\newpage

\section{Introduction}
Universities continue to rapidly evolve to address the needs of diverse, growing student populations, while embracing advances in pedagogy and technology. With the revolution in big data, universities are striving to utilise their rich and complex digital data on learners towards the personalisation of education.

At The University of Queensland, we have developed an adaptive, student-facing learning platform
called RiPPLE (Recommendation in Personalised Peer Learning Environments) that provides
personalised content and learning support at scale. RiPPLE employs exemplary techniques from the fields of machine learning, crowdsourcing, learning analytics and recommender
to provide the following main functionalities:
\begin{itemize}
\item \textit{Co-creation}: \name empowers learners to contribute to co-creation of learning content.
\item \textit{Knowledge tracing}: \name approximates learners' knowledge states based on their interactions with the platform. It uses a variety of visualisations to enable learners to track their knowledge states and progress.
\item \textit{Content recommendation}: \name uses the knowledge state of learners to recommend formative exercises tailored to the needs of each individual.
\item \textit{Peer Learning Support}: \name recommends peer learning sessions based on the availability, learning preferences and needs of individuals. Students are given the opportunity to provide peer learning support, seek peer learning support or find study partners. 
\item \textit{Gamification}: \name uses leaderboards and a badging system to increase student motivation and performance.
\end{itemize}

The RiPPLE platform has been released as an open-source package\footnote{\url{https://github.com/hkhosrav/RiPPLE-Core/}} under the Apache 2.0 license and is freely available for usage in non-commercial settings. A prototype of the system is accessible through a GitHub account\footnote{\url{https://hkhosrav.github.io/RiPPLE-Core/\#/}}. RiPPLE can use the Learning Tools Interoperability (LTI) standard to integrated into many popular learning management systems including Blackboard, Moodle and Canvas. It is designed to be scalable, allowing the system to be easily adopted by many courses across different faculties in many universities. It is also designed to be sustainable, requiring very low maintenance from the teaching team. 

The rest of this paper is organised as follows: Section~\ref{sec:relatedWork} provides an overview of the related work on knowledge tracing, adaptive learning, recommender systems in technology enhanced learning (RecSysTEL), and reciprocal recommendation. Section~\ref{sec:notation} describes the data sources, functionality, the main aims and outputs of \name using a formal notation, which can be used by educational researchers. Section~\ref{sec:platform} presents an overview of the main features and functionalities that are supported by \name including creating and answering questions, knowledge tracing and recommending questions, reciprocal peer recommendation, leaderboards, personal profiles, and a page designed for instructors. Section~\ref{sec:benefits} discusses the expected benefits students, instructors, and educational researchers will receive from using \name. Section~\ref{sec:client} provides information on the implementation of the \name Client and Section~\ref{sec:server} provides information on the implementation of the \name Server. Finally, Section~\ref{sec:contributing} provides guidelines around contribution to \name.


\section{Related Work}\label{sec:relatedWork}

\subsection{Knowledge Tracing}
Knowledge tracing is the task of modelling the knowledge state of students so that their future performance on learning activities can accurately be predicted \cite{corbett1994knowledge}. The problem of knowledge tracing was introduced, and has been heavily studied within the intelligent tutoring community \cite{corbett2001cognitive}. The Bayesian Knowledge tracing (BKT) algorithm \cite{corbett1994knowledge} is one of the most prominent methods used for knowledge tracing. BKT uses Hidden Markov Models to capture the student knowledge states as a set of binary variables representing whether or not a concept has been mastered. 

BKT has recieved significant attention and improvement since it was first proposed. \citeN{Baker2008} introduced slipping and guessing parameters; slipping refers to the situation where a student has the required skill for answering a question but mistakenly provides the wrong answer; and guessing refers to the situation where a student provides the right answer despite not having the required skill for solving the problem. Later on, \citeN{pardos2011kt} effectively extended BKT to capture item difficulty, which led to improve prediction accuracy. More recently, \citeN{yudelson2013individualized} further improved BKT by introducing a new set of parameters capturing prior knowledge of individual learners. 

Other algorithms with comparable or superior predictive power to BKT have also been proposed for knowledge tracing. \citeN{cen2006learning} introduced the Learning Factors Analysis Framework, and \citeN{pavlik2009performance} introduced the Performance Factor Analysis framework. \citeN{khajah2014integrating} incorporated Item Response theory (IRT) into knowledge tracing, and more recently, \citeN{Piech2015} and \citeN{sha2017neural} used recurrent neural networks for deep knowledge tracing.

The current implementation of \name uses the knowledge tracing algorithm of \citeN{khosravi2017}
\subsection{Adaptive Learning}

Adaptive learning platforms use knowledge tracing to dynamically adjust the level or type of instruction based on individual student abilities or preferences \cite{paramythis2003adaptive}. Use of adaptive platforms help personalise learning, and has been shown to improve or accelerate student performance \cite{yilmaz2017effects}. At a high level of generality all adaptive learning platforms rely on four interacting models: (1) a knowledge space modelling what the learners need to know, (2) a set of knowledge states modelling what students currently know, (3) a repository of learning objects, mapped to the concepts of the knowledge space, modelling the learning activities that are available to the students, (4) a recommender system modelling the extend to which different learning activities meet the learning needs of each of the students \cite{essa2016possible}.  

There are two main types of adaptive learning platforms referred to as the publisher model and the platform model \cite{oxman2014white}. In the publisher model, the system is designed with pre-existing content, often based on textbooks from the publisher. Pearson’s MyLabs (using Knewton \cite{Ferreira2016white} for its adaptive functionality), McGraw-Hill’s LearnSmart and ALEKS \cite{falmagne2006assessment} are established examples of this model.  The publisher model has been very successful in K-12, where course content often have to comply with national standards. However, the overall adoption rate of adaptive learners that use the publisher model in higher education has been very low, and has been mostly restricted to research projects 
\cite{essa2016possible}. Limitations in tailoring the course content and the high per student cost of using these systems are among the main factors that contribute to the low adoption rate.

The platform model provides a content-agnostic system infrastructure that enables the teaching team to develop  and author the content of their course. Smart Sparrow \cite{Sparrow2016} and many learning management systems such as Desire2Learn, Loudcloud and edX that incorporate adaptive functionality into their course building tools follow this model. The platform model is relatively new and mostly suffers from an operational limitation rather than a technological one; implementing adaptivity in a course requires a large amount of new content and object tagging, which introduces a significant overhead for the teaching time. \name uses crowdsourcing to overcome this challenge, lowering the required workload associated with the adoption of adaptive learning using the platform model.

\subsection{Recommender Systems in Technology Enhanced Learning}

RecSysTEL is an active and rapidly evolving research field. For example, \citeN{Drachsler2015} perform an extensive classification of 82 different RecSysTEL environments, and \citeN{Erdt2015} review the various evaluation strategies that have been applied in the field. Together these articles provide recent comprehensive surveys that consider more than 200 articles spanning over 15 years. 
Collaborative filtering (CF) identifies similar users and provides recommendations based upon their usage patterns. CF has been extensively employed in RecSysTEL; an early LAK paper, \citeN{Verbert2011} evaluated and compared the performance of different CF techniques on educational data sets, showing that the best choice of algorithm is data dependent. In a more recent study, \protect\citeN{Kopeinik2017}
also concluded that the performance of the algorithms strongly depends on the properties and characteristics of the particular dataset.
In combining educational data sets with social networks, \citeN{Cechinel2013} used CF to predict the utility of items for users based on their interest and the interest of the network of the users around them. Similarly, \citeN{Fazeli2014} proposed a graph-based approach that uses graph-walking for improving performance on educational data sets. 


One important way in which RecSysTEL has been used in an educational setting is to recommend personalized learning objects. Thus, \citeN{Lemire2005} used inference rules to provide context aware recommendation on learning objects, and \citeN{Mangina2008} recommend documents and resources within e-learning environments to expand or reinforce knowledge. Interestingly, \citeN{Gomez-Albarran2009} combined content based filtering with collaborative filtering to make recommendations in a student authored repository. When recommending learning objects (e.g. questions) related to student, \citeN{Cazella2010} provided a semi-automated, hybrid solution based on CF (nearest neighbor) and rule based filtering, while \citeN{Thai-Nghe2011} used students' performance predictions to recommend more appropriate exercises. CF techniques (basic, biased, and tensor matrix factorization) were used to address a number of different student behaviors and to model the temporal effect of students improving over time. Recently, \citeN{Imran2016} provided an automated solution to personalize a learning management systems (LMS) using advanced learners' profiles to encapsulate their expertise level, prior knowledge, and performance in the course. The approach used association rule mining to create the learning object recommendations. 

Matrix factorization (MF) is one of the most established techniques used in CF; however, despite its success in RecSys, MF has rarely been used in RecSysTEL. Out of the 124 papers referenced by \citeN{Drachsler2015}, only two papers directly use it \cite{Salehi2013,Thai-Nghe2011}. This is somewhat surprising; MF has been put to good use in EDM for generating latent profiles of student expertise and so ought to combine with RecSysTEL in a straightforward manner. Indeed, the intelligent tutoring systems that appear to be utilized in that body of work could be seen as closely related to RecSysTEL, although they tend to give students less autonomy to accept or reject the pathways chosen for them \cite{chen2008}. MF is particularly powerful in modeling students' performance and knowledge, because it implicitly incorporates guess and slip factors as latent factors \cite{Thai-Nghe2011}. The current implementation of \name uses the recommender system of \citeN{khosravi2017}, which employs MF.

\subsection{Reciprocal Recommendation}
Reciprocal recommender systems has a literature rich with applications in online dating \cite{pizzato2013},  job matching \cite{Hong2013}, and social networking \cite{Guy2015}, each highly tailored by the best consideration of the specific needs and constraints of the immediate environment.  
Fundamentally they entail operationalised user preferences and seek solutions such as a list of recommendations that match those preferences to a higher degree than competing items. In reciprocal recommendation, items are usually other users whose preferences must also be fulfilled and so entail a higher level of complexity than other recommender systems \cite{pizzato2010reciprocal}. Much of the research in social recommending has been developed and evaluated in existing social networks and particularly online dating sites where the site and/or the individual users may have a substantial data history which can be used to train machine learning algorithms \cite{cai2011learning,chen2013,kutty2014}.  In peer recommending the environment may be a newly formed cohort of learners (such as a first-year university course) who have no data history but can volunteer preferences on a narrow selection of relevant variables for the purposes of matching.

By their nature, recommender systems across domains will exhibit similarities to a high degree. All systems incontrovertibly share the same fundamental goal - to provide recommendations that are well-received by users according to their preferences, explicit and implicit, in an otherwise overwhelming information environment where the likelihood of users successfully finding preferred items without technological assistance is very low. However the nature of domain-specific information and definition of a successful recommendation is so heavily context- and goal-dependent that little more than the general way of thinking can be adapted or generalised from existing systems to new domains. This is particularly true of the formulation of user preference models upon which recommendations are to be based, making them necessarily bespoke. 


As such, while reciprocal peer recommendation has similarities to traditional recommendations in education and reciprocal recommendations in other domains, it does have a distinct nature. Although some primary research has been done on utilising peer learning and support for improving learning and enhancing the learning experience of students, the area remains fertile for many research and development opportunities. In this paper we introduce a platform that can enable adoption of peer support systems for both large on-campus and online courses with competency-based user preference models.   

The current implementation of \name uses the reciprocal peer recommendation algorithm of \cite{potts2018reciprocal}.

\section{Formal Notation}\label{sec:notation}
This section describes the data sources, functionality, the main aims and outputs of \name using a formal notation. 

\paragraph{Users and Questions:} Let  $U_N =\{ u_1 \ldots u_N \}$ denote the set of users that are enrolled in a course in \name, where $u_n$ and $u_{n'}$ refer to arbitrary users. 
Let $ Q_M = \{ q_1 \ldots q_M \}$ denote the repository of the questions that are available to users in a course in \name, where $q_m$ and $q_{m'}$ refer to arbitrary questions. All of the events occurring in \name are logged using a set of timestamps $T_T = \{ \tau_1 \ldots \tau_T \}$, where $\tau_t$ refers to an arbitrary timestamp. A three-dimensional array $C_{N\times M}^T$ keeps track of question creations, where $c_{nm}^t$ indicates that user $u_n$ has created question $q_m$ at timestamp $\tau_t$. Similarly, a three-dimensional array $A_{N\times M}^T$ keeps track of question answers, where $a_{nm}^t=1$ indicates that user $u_n$ has answered question $q_m$ correctly at timestamp $\tau_t$, and $a_{nm}^t=0$ indicates that user $u_n$ has answered question $q_m$ incorrectly at timestamp $\tau_t$. In addition, a three-dimensional array $D_{N\times M}^T$ provides information on question difficulty perceptions, where $d_{nm}^t$ is the difficulty level user $u_n$ has expressed for question $q_m$ at timestamp $\tau_t$. Furthermore, a three-dimensional array $R_{N\times M}^T$  provides information on question ratings, where $r_{nm}^t$ is the rating user $u_n$ has expressed for question $q_m$ at timestamp $\tau_t$. 

\paragraph{Knowledge Tracing and Question Recommendation:} Each course consists of a set of knowledge units $\Delta_L = \{ \delta_1 \ldots \delta_L \}$ referred to as a knowledge space, where $\delta_l$ refers to an arbitrary knowledge unit. Questions can be tagged with one or more knowledge units;  $\Omega_{M\times L}$ is a two-dimensional array, where $\omega_{ml}$ is  $\frac{1}{g}$ if question $q_m$ is tagged with $g$ knowledge units, including  $\delta_l$ and 0 otherwise. One of the initial aims of \name is to use knowledge tracing algorithms to approximate the knowledge state of students on each knowledge unit. A three-dimensional array $\Lambda_{N \times L}^T$ is used for representing students' knowledge states approximated by the system, where $\lambda_{nl}^t$ represents the knowledge state of $u_n$ on $\delta_l$ at timestamp $\tau_t$. This information is used to produce a three dimensional-array $\Theta_{N\times M}^T$, where  $\theta_{nm}^t$ shows the personalised score of question $q_m$ for user $u_n$ at timestamp $\tau_t$. $\Lambda_{N \times L}^T$ can be used for recommending questions to each user. 

\paragraph{Reciprocal Peer Recommendation}: Let $V_J= \{ v_1 \ldots v_J \}$ indicate a set of time slots, which  denote the weekly available time slots for scheduling a study session, where $v_j$ refers to an arbitrary time slot. Students can hold different roles $E_3 = \{ e_1, e_2, e_3 \}$ for participation in peer learning sessions, where $e_1$ is used to present providing peer learning support, $e_2$ for seeking peer learning support, and $e_3$ for searching for study partners. $e_o$ refers to an arbitrary role. A four-dimensional array $S_{N \times L \times 3}^T$ represents the requests of the students, where $s_{nlo}^t=1$ indicates that user $u_n$ has indicated interest in participating in a study session on knowledge unit $\delta_l$ with role $e_o$ at timestamp $\tau_t$. In addition, a three-dimensional array $Z_{U \times V}^T$ represents the availability of students, where $z_{nj}^t$ shows the availability of user $u_n$ for time slot $v_j$ at timestamp $\tau_t$. Furthermore, A three-dimensional array $P_{N \times 3}^T$ shows competency preference of students in study sessions, where $P_{no}^t$ shows the competency preference of user $u_n$ for roles $e_o$ at timestamp $\tau_t$. For example,  $P_{n1}^t = \epsilon$ means that at timestamp $\tau_t$, $u_n$ prefers providing support to peers with a competency of around $\epsilon$ less than his own. To be able to provide meaningful recommendations, we constrain eligibility by role such that users (1) provide support to less competent learners, (2) seek support from more competent learners and (3) find study partners with relatively similar competency to that of their own. Using all of this information, \name aims to compute a three-dimensional array $\Psi_{N\times N}^T$, where $\psi_{nn'}^t$ shows the reciprocal score between user $u_n$ and user $u_{n'}$ at timestamp $\tau_t$. $\Psi_{N\times N}^T$ can be used for recommending peers to one another.

Table~\ref{tab:notation} provides a summary of the notation used for describing the functionality of \name.
\begin{table}[ht]
\centering
\footnotesize{
\begin{tabular}{|p{1.2cm}|p{15cm}|}
\hline
\multicolumn{2}{|c|}{Users and Questions}  \\ 
\hline

$U_N$       & A set of users $\{ u_1 \ldots u_N \}$, where $u_n$ and $u_{n'}$ refer to arbitrary users.  \\
$Q_M$		  &A set of questions $\{ q_1 \ldots q_M \}$, where $q_m$ and $q_{m'}$ refer to arbitrary questions.\\
$T_T$ & A set of time stamps $\{ \tau_1 \ldots \tau_T \}$, where $\tau_t$ refers to an arbitrary timestamp. \\
$C_{N\times M}^T$       & A three-dimensional array where $c_{nm}^t$ indicates that user $u_n$ has created question $q_m$ at timestamp $\tau_t$.  \\
$A_{N\times M}^T$       & A three-dimensional array where $a_{nm}^t=1$ indicates that user $u_n$ has answered question $q_m$ correctly at timestamp $\tau_t$ and 0 if answered incorrectly.  \\
$D_{N\times M}^T$       &A three-dimensional array where $d_{nm}^t$ is the difficulty level user $u_n$ has expressed for question $q_m$ at timestamp $\tau_t$. \\
$R_{N\times M}^T$       &A three-dimensional array, where $r_{nm}^t$ is the rating user $u_n$ has expressed for question $q_m$ at timestamp $\tau_t$. \\   
\hline
\multicolumn{2}{|c|}{Knowledge Tracing and Question Recommendation}  \\ 
\hline
$\Delta_L$   & A set of knowledge units referred to as knowledge space $\{ \delta_1 \ldots \delta_L \}$ where $\delta_l$ refers to an arbitrary  knowledge unit.  \\

$\Omega_{M\times L}$       &A  matrix, where $\omega_{ml}$ is  $\frac{1}{g}$ if question $q_m$ is tagged with $g$ knowledge units, including  $\delta_l$ and 0 otherwise.\\ 
$\Lambda_{N \times L}^T$       & A three-dimensional array, where $\lambda_{nl}^t$ represents the knowledge state of $u_n$ on $\delta_l$ at timestamp $\tau_t$.\\
$\Theta_{N\times M}^T$       & A three-dimensional array, where $\theta_{nm}^t$ shows the personalised score of question $q_m$ for user $u_n$ at timestamp $\tau_t$.\\

\hline
\multicolumn{2}{|c|}{Reciprocal Peer Recommendation}  \\ 
\hline
\\ 
$V_J$      &  A set of time slots $\{ v_1 \ldots v_J \}$, which  denote the weekly available time slots for scheduling a study session, where $v_j$ refers to an arbitrary time slot.  \\

$E_3$ & A set of roles $\{ e_1, e_2, e_3 \}$ for participation in peer learning sessions where $e_1$ is used to present providing peer learning support, $e_2$ for seeking peer learning support, and $e_3$ for searching for study partners. $e_o$ refers to an arbitrary role. \\

$S_{N \times L \times 3}^T$ & A four-dimensional array, where $S_{nlo}^t=1$ indicates that user $u_n$ has indicated interest in participating in a study session on knowledge unit $\delta_l$ with role $e_o$ at timestamp $\tau_t$.\\

$Z_{U \times V}^T$ & A three-dimensional array in which $z_{nj}^t$ shows the availability of user $u_n$ for time slot $v_j$ at timestamp $\tau_t$.\\ 
$P_{N \times 3}^T$ & A three-dimensional array in which $P_{no}^t$ shows the competency preference of user $u_n$ for roles $e_o$ at timestamp $\tau_t$.\\ 
$\Psi_{N\times N}^T$       & A three-dimensional array, where $\psi_{nn'}^t$ shows the reciprocal score between user $u_n$ and user $u_{n'}$ at timestamp $\tau_t$.\\     
\hline         
\end{tabular}
}
\caption{A summary of the notation used for describing the functionality of \name}
\label{tab:notation}
\end{table}

\section{The RiPPLE Platform}\label{sec:platform}
This section presents an overview of the main features and functionalities that are supported by \name.

\subsection{Creating and Answering Questions}
\begin {figure*}[h!]
\centering
\includegraphics[width=17.4 cm, height=10 cm]{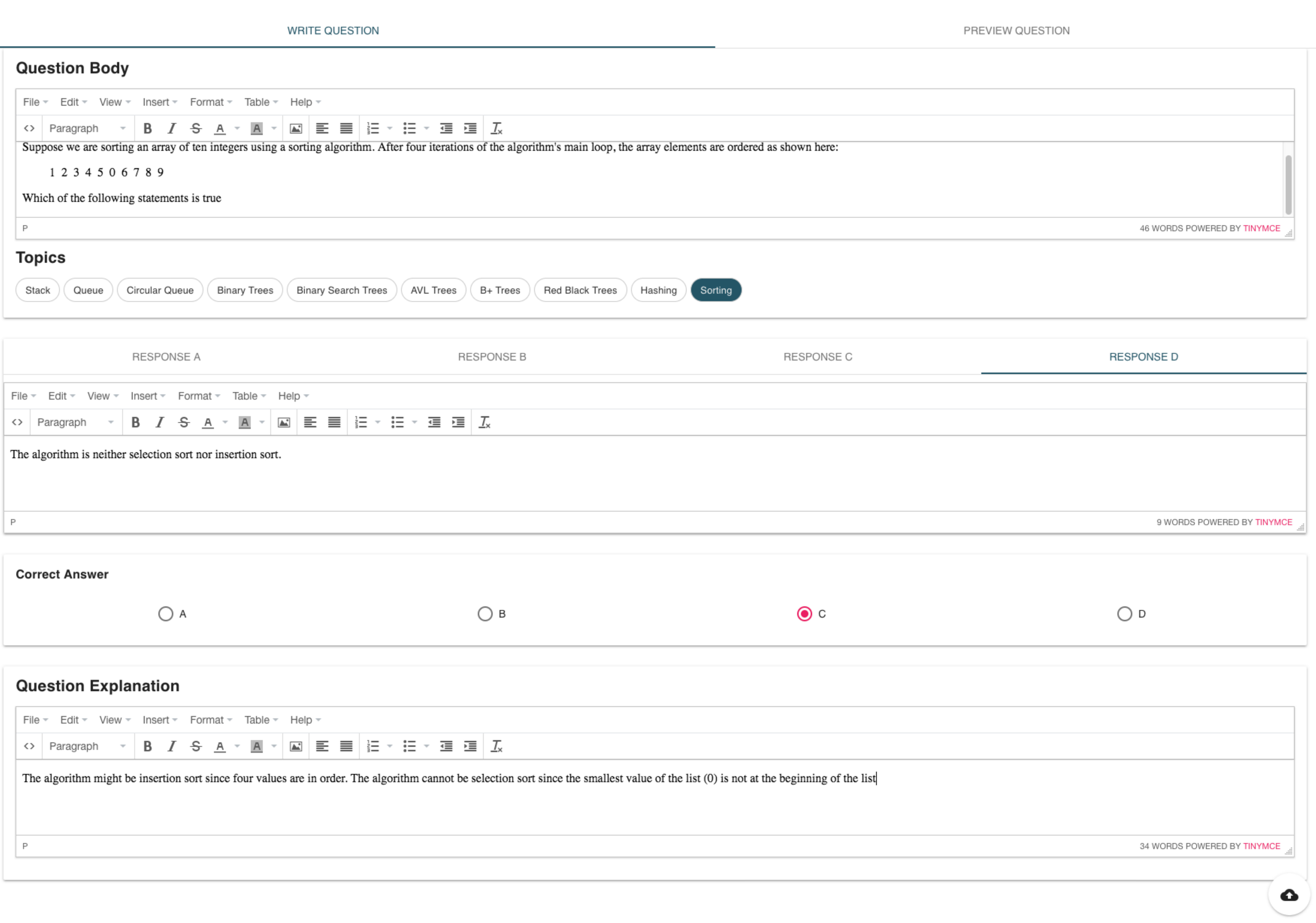}
\caption{Overview of the question authoring page in \name.\label{fig:questionauthoring}} 	
\end {figure*}

There are many benefits in engaging students as partners in co-creation of educational content \cite{bovill2011students}, and in particular, in creation of multiple-choice questions \cite{Denny2008}. \name enables students to create questions and share them with their peers. Figure~\ref{fig:questionauthoring} shows the graphical interface used in \name for creating questions. 

Creating a question includes the following steps:
\begin{enumerate}
\item Developing the body of the question. Text, images, videos and scientific formulas may be used in the development of the body of the question.
\item Tagging the question with one to four topics. The available topics are pre-defined by the instructor of the course.
\item Authoring the multiple-choice answers. Similar to the body of the question, Text, images, videos and scientific formulas may be used in the development of each of the multiple-choice answers.
\item Nominating the correct answer and developing a solution. An ideal solution includes rationale for the correctness of the right multiple-choice answer and the lack of correctness of the other multiple-choice answers. 
\item previewing the question to make sure that it renders correctly and as expected.
\item submitting the question to be stored as part of the question repository. 
\end{enumerate}

\begin {figure*}[h!]
\centering
\includegraphics[width=17.4 cm, height=10 cm]{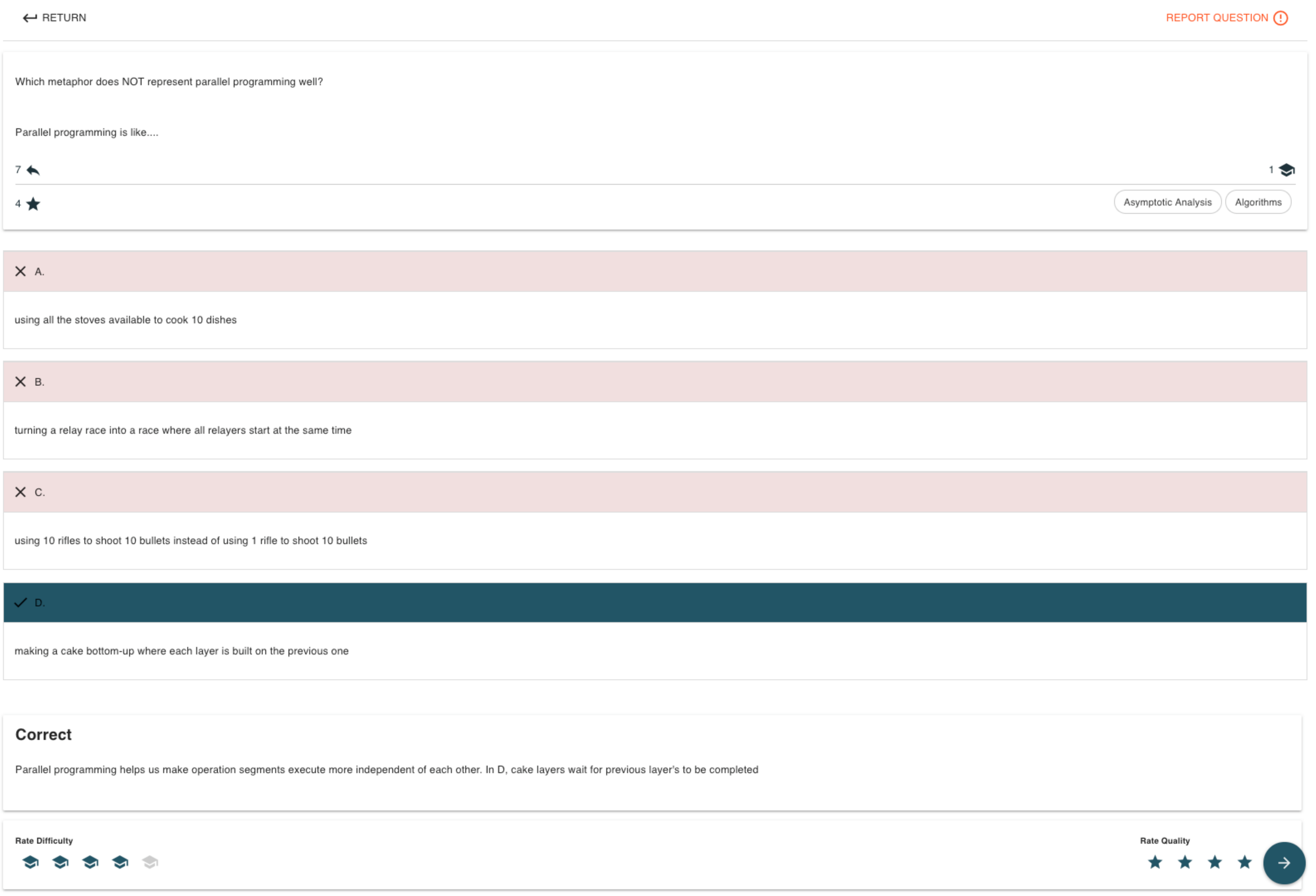}
\caption{Overview of the question answering page in \name. \label{fig:answerQuestion}} 	
\end {figure*}

Students are able to answer and rate questions that are available on the platform. Once they have answered a question, they are able to view the right answer, the distribution of how the question has been answered by their peers, and the explanation provided for the answer. Students are then able to rate the quality and the difficulty of the question. Figure~\ref{fig:answerQuestion} shows the graphical interface used in \name for answering questions. 

\begin {figure*}[h!]
\centering
\fbox{
\includegraphics[width=5.4 cm]{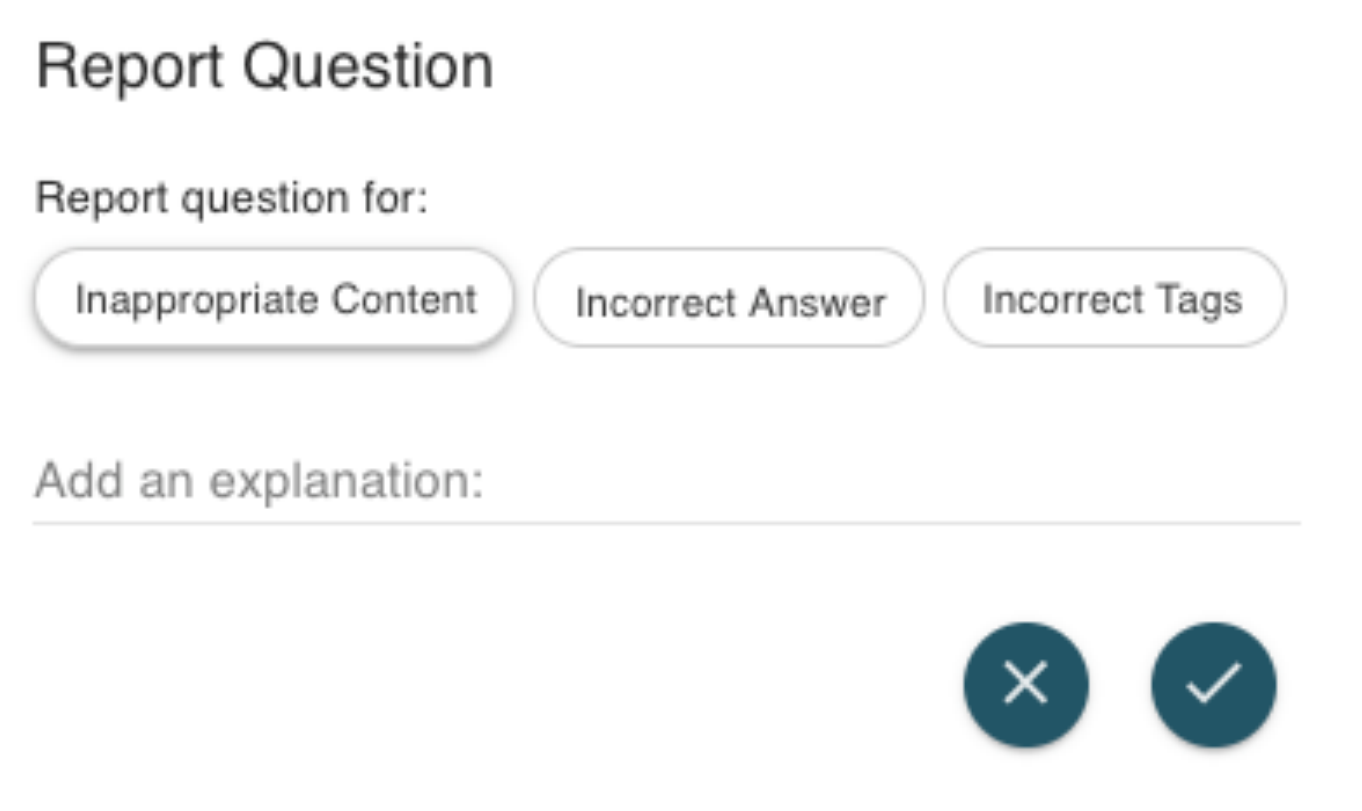}
}
\caption{The interface used for flagging and reporting inappropriate questions in \name.\label{fig:reportQuestion}} 	
\end {figure*}

\name also relies on crowdsourcing to identify inappropriate or incorrect questions. Figure~\ref{fig:reportQuestion} shows how questions may be flagged as inappropriate in \name.
Users with the ``instructor" role have the ability to view, edit and delete questions that are flagged as inappropriate.

\subsection{Knowledge Tracing and Recommending Questions}
\begin {figure*}[h!]
\centering
\includegraphics[width=17.4 cm, height=10 cm]{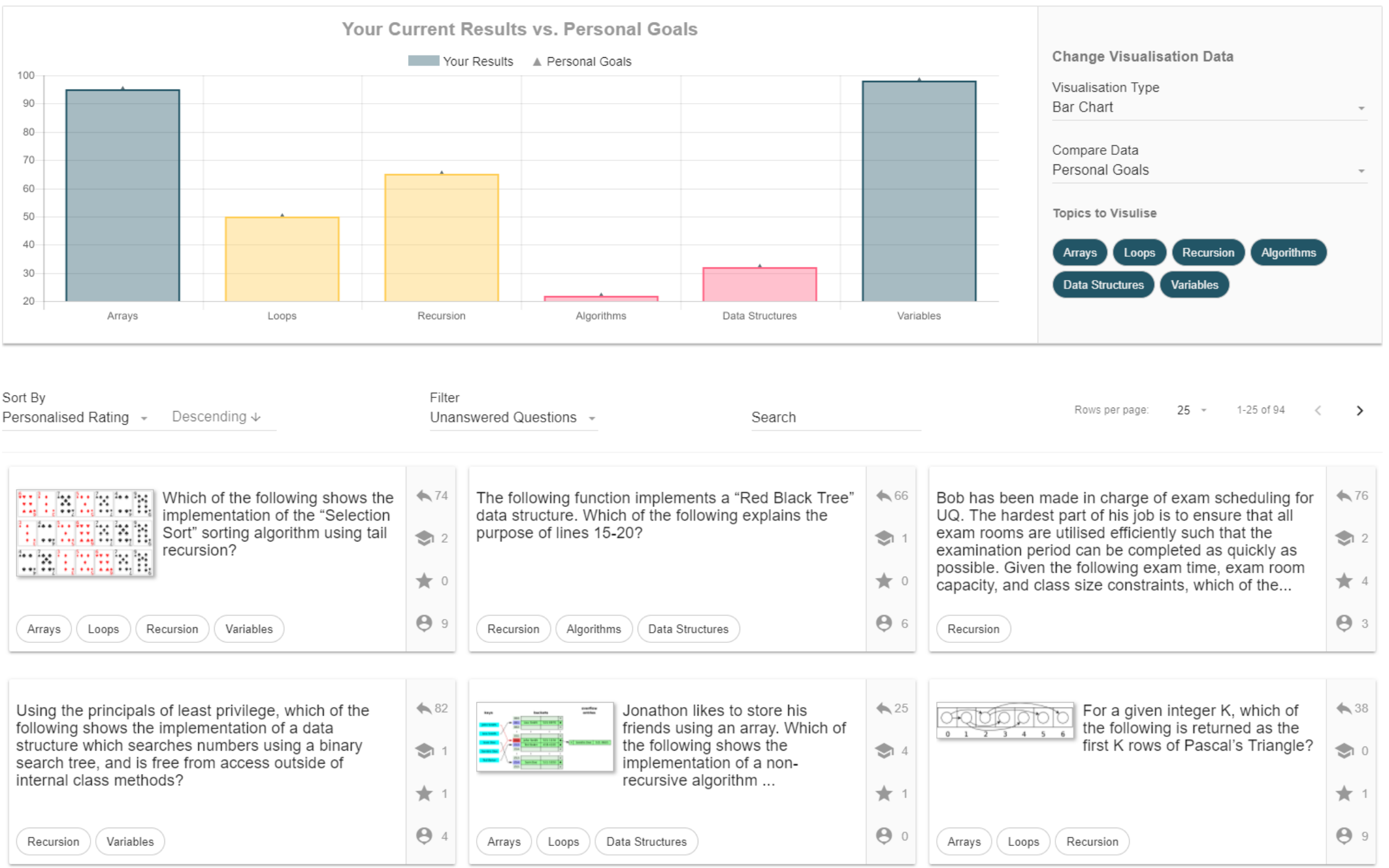}
\caption{Overview of the knowledge tracing and question recommendation page of \name.\label{fig:Dashboardoverview}} 	
\end {figure*}

One of the main pages of \name, as shown in Figure~\ref{fig:Dashboardoverview}, is dedicated to knowledge tracing and recommending questions. The top section of this page provides an interactive visualisation widget that enable learners to select their desired visualisation type for viewing their knowledge state. 

``Visualisation Type'' allows learners to select from a range of different visualisation techniques so that they can choose a visual display that better suits their comprehension and personal preference. \name is equipped with a set of different types of visualisations: Bar charts (as displayed in Figure~\ref{fig:Dashboardoverview}) are helpful in presenting rich information regarding users. They are simple to read and are comprehensible by a wide audience. Colour of the charts categorises competencies into three levels: red demonstrates inadequate competency in a topic, yellow demonstrates adequate competency with room for improvement, and blue demonstrates mastery in a topic. Radar charts are visually striking, and can add interest to what would otherwise be a dry data presentation. One of the strengths of radar charts is that they support visualisation of multiple variables consisting of measures that require different quantitative scales, which a bar chart cannot accommodate. They are very well suited for comparing the knowledge states of users students. 
Box plots are an effective way of displaying the distribution of data based on minimum, median, maximum, 1st quartile and 3rd quartile. Box plot displays may be preferred when displaying data of a group of learners as they determine if the data is skewed based on where the median sits within the box relative to the inter-quartile ranges. However, they are harder than both bar charts and radar charts to read and comprehend.
The widget also supports the use of more recently developed visualisations that are tailored towards education. For example, it supports the use of Topic Dependency Models \cite{Cooper2018}, which use two-weighted graphs to display the knowledge state of learners not only on based on individual topics but also on combination of two or more topics.

``Compare Data'' allows learners to compare their knowledge states against a range of options: 
``Peers'' mode enables them to compare their performance with a selected distribution (e.g. top 20\%) of the peers that are currently enrolled in the course. ''Previous Offerings'' enables learners to compare their performance with a selected distribution of learners across all offerings of the course. ``Topic to Visualise'' option enables users to select topics in which their competencies are visualised.

The bottom section of this page, as shown in Figure~\ref{fig:Dashboardoverview}, enables learners to select questions using search and recommendation functionalities. The ``Sort By'' option allows learners to sort questions based on their difficulty, quality, number of responses, number of comments or personalised rating. By selecting "Personalised Rating", the platform sorts the questions based on the outcome of recommender system. The ''Filter'' option enables users to filter the questions that are included in the results. They can request all questions (default), unanswered questions, answered questions, or wrongly answered questions to be included in the results.  The ``Search'' option enables learners to search for questions based on specific content that may be present in the questions or multiple choice answers.

\begin {figure}[h!]
\centering
\includegraphics[width=8.5 cm]{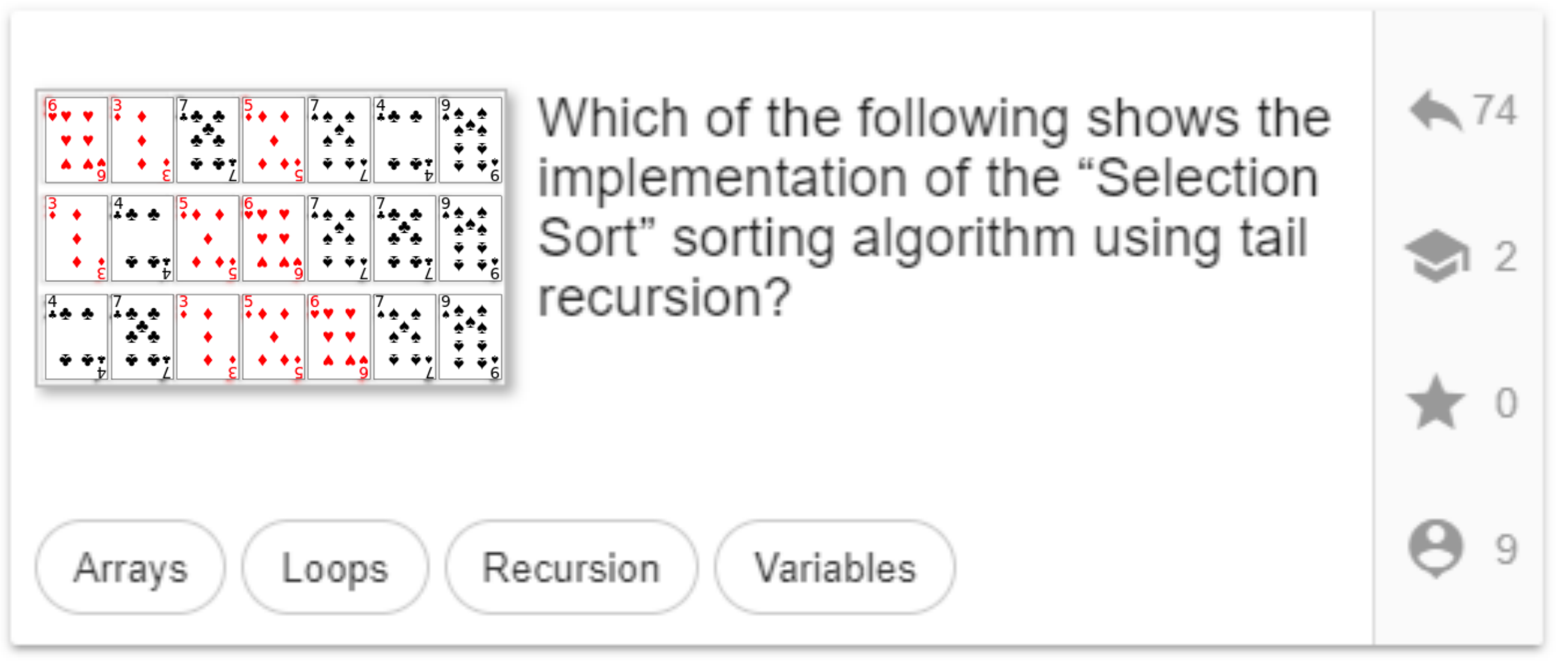}
\caption{A sample question card in \name.\label{fig:questioncard}}
\end {figure}

The results of the search are presented as a list of question cards, allowing users to engage with questions that best suit their needs. Figure~\ref{fig:questioncard} shows a sample question card. Each question card includes an overview of the question content, the topics associated with the question, and a sidebar in which the first icon shows the number of responses to the question, the second icon shows the average difficulty rating of the question, the third icon shows the average quality rating of the question, and the last icon shows the personalised rating that demonstrates the suitability of the question for each learner. Clicking on the question card will take the users to another page that would allow them to answer and rate the question.


\subsection{Reciprocal Peer Recommendation}
\begin {figure*}[h!]
\centering
\includegraphics[width=17.4 cm]{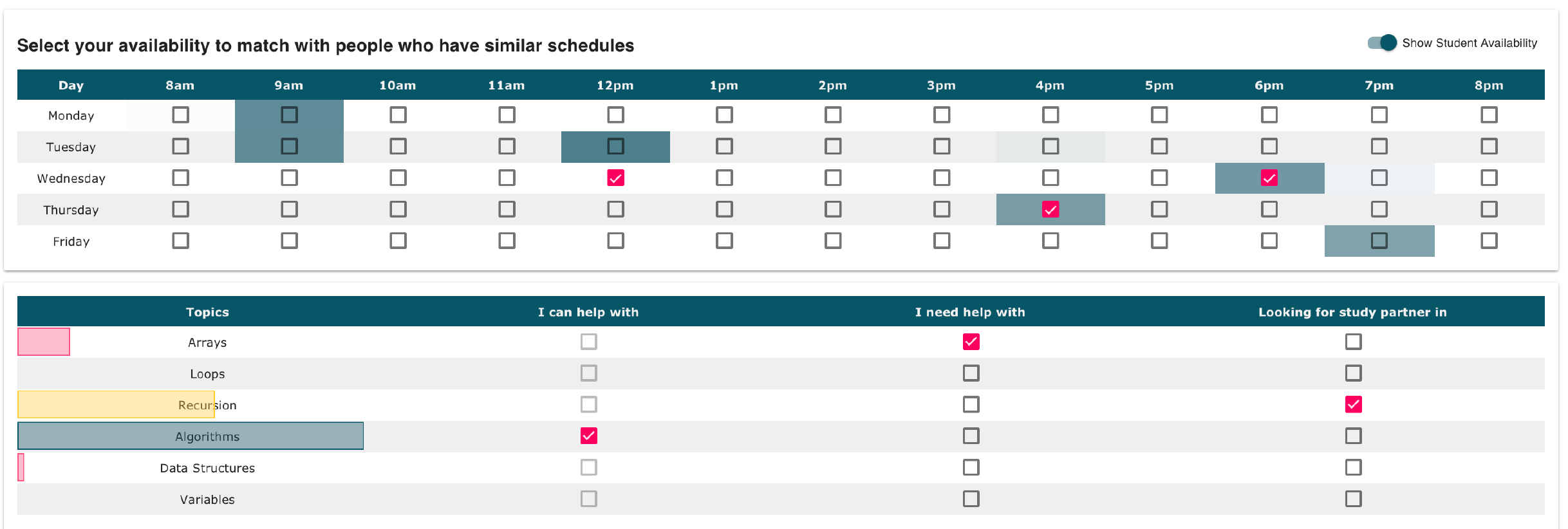}
\caption{Overview of the peer recommendation interface in \name.\label{fig:platformoverview}}
\end {figure*}

One of the main features of \name is its ability to recommend peer learning sessions.
Learners nominate their availability in hourly blocks, and their preferences for providing or seeking peer learning support and finding study partners across the range of course-relevant topics. Figure~\ref{fig:platformoverview} shows the graphical interface used for capturing this information in \name.

The shade around each of the time slots provides an indication of the popularity of that time, where darker shades indicate a higher amount of interest in the time slot. The knowledge state of the student is provided in the form of a coloured bar chart superimposed over the list of topics. The option to provide peer support is only available for those topics in which the student meets a required competency threshold, denoted in the bar chart by the colour blue. The knowledge state of students are updated using the their cumulative performance on assessment items progressively during the teaching period using algorithms described by in \citeN{khosravi2017}.




\begin {figure}[h!]

\centering
\includegraphics[width=7 cm]{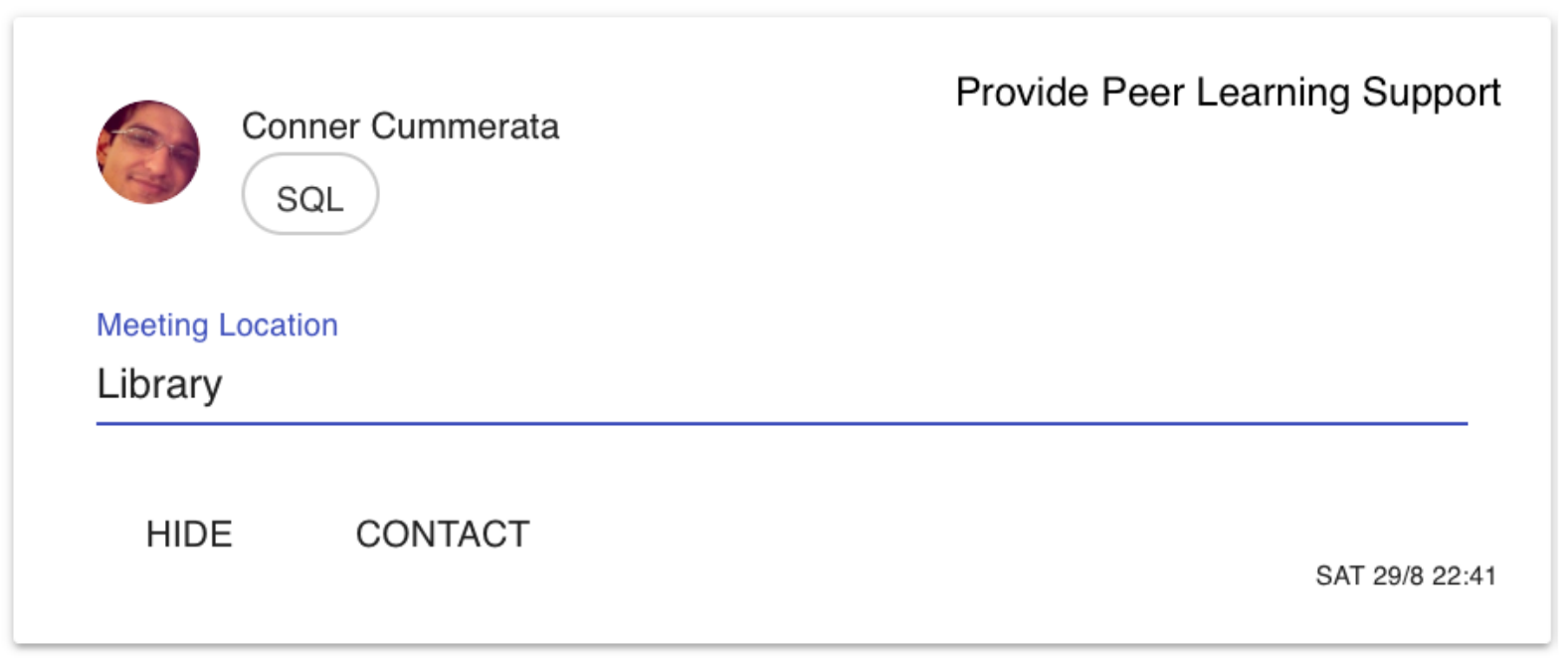}
\caption{Example recommendation for a learner to provide peer learning support.\label{fig:recommendation}}
\end {figure}

Figure~\ref{fig:recommendation} provides an example of a peer support recommendation in the student interface, identifying the potential peer learning supporter and the two topics for which support is available. Students then have the option to ignore or request a meeting with the recommended peer at the nominated time and date.

\subsection{Leaderboard}
\begin {figure*}[h!]
\centering
\includegraphics[width=17.4 cm]{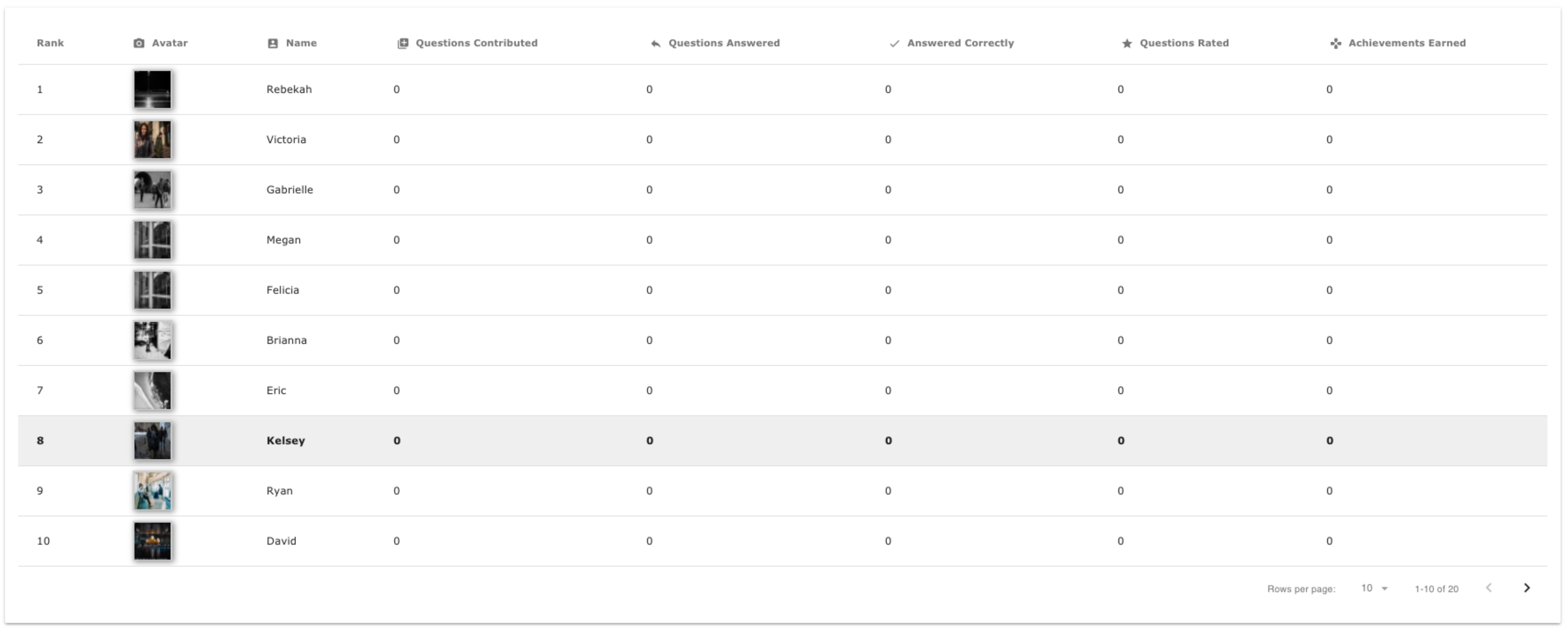}
\caption{Overview of the leaderboard in \name \label{fig:leadersboard}}
\end {figure*}

A “leaderboard”, as its name implies, displays individuals with highest score on a give task. Use of leaderboards in education has shown to increase student motivation and engagement \cite{landers2014empirical,banfield2014increasing}. The leaderboard in \name displays students with the highest score on a variety of items including the number of questions contributed, answered, correctly answered, and rated. It also displays the students with the highest number of achievements, which are presented in terms of gamified badges. Figure~\ref{fig:leadersboard} shows the leaderboard used in \name.

\subsection{Personal Profile}
\begin {figure*}[h!]
\centering
\includegraphics[width=10.4 cm]{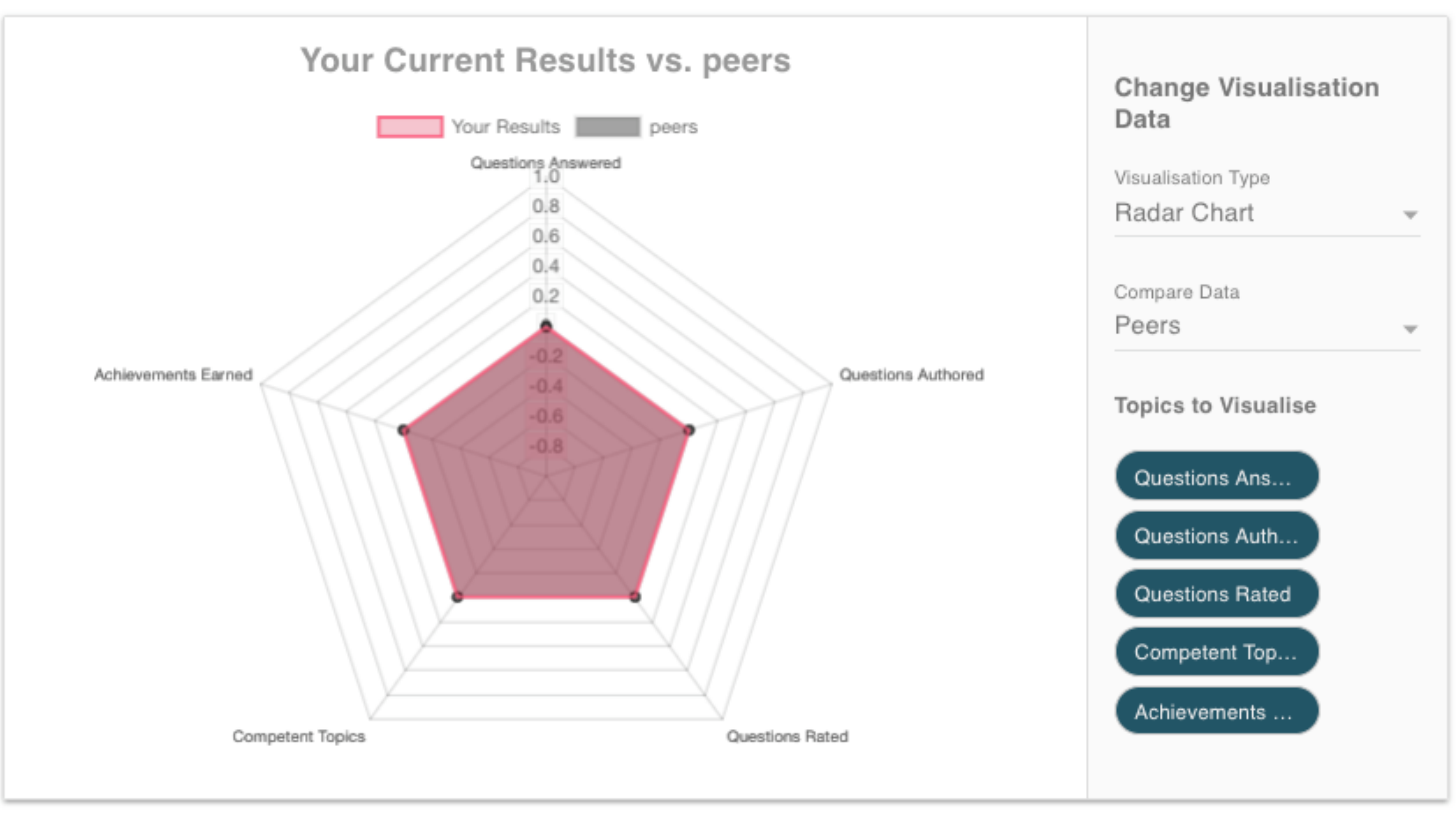}
\caption{The visualisation widget used for showing engagement in \name \label{fig:engagementVis}}
\end {figure*}

Each student is provided with a personal profile that includes information on their engagement, achievements, notifications, and their consent on use of their data for educational research purposes. 
\paragraph{Engagement}
The engagement level of students on a variety of tasks are presented using a visualisation widget. This widget enables students to compare their engagement using a visualisation type of their choice against their peers or their own targeted goals on a set of tasks. The default visualisation type uses Kiviat diagrams, which are more informally known as radar charts. Kiviat diagrams have been used extensively in visualising educational dashboard (e.g., see \cite{may2011travis}) as they are able to display multivariate observations with an arbitrary number of variables \cite{chambers1983graphical}. The visualisation widget Figure~\ref{fig:engagementVis} shows the chart used for presenting engagement in \name. 

\begin {figure*}[h!]
\centering
\includegraphics[width=14.4 cm]{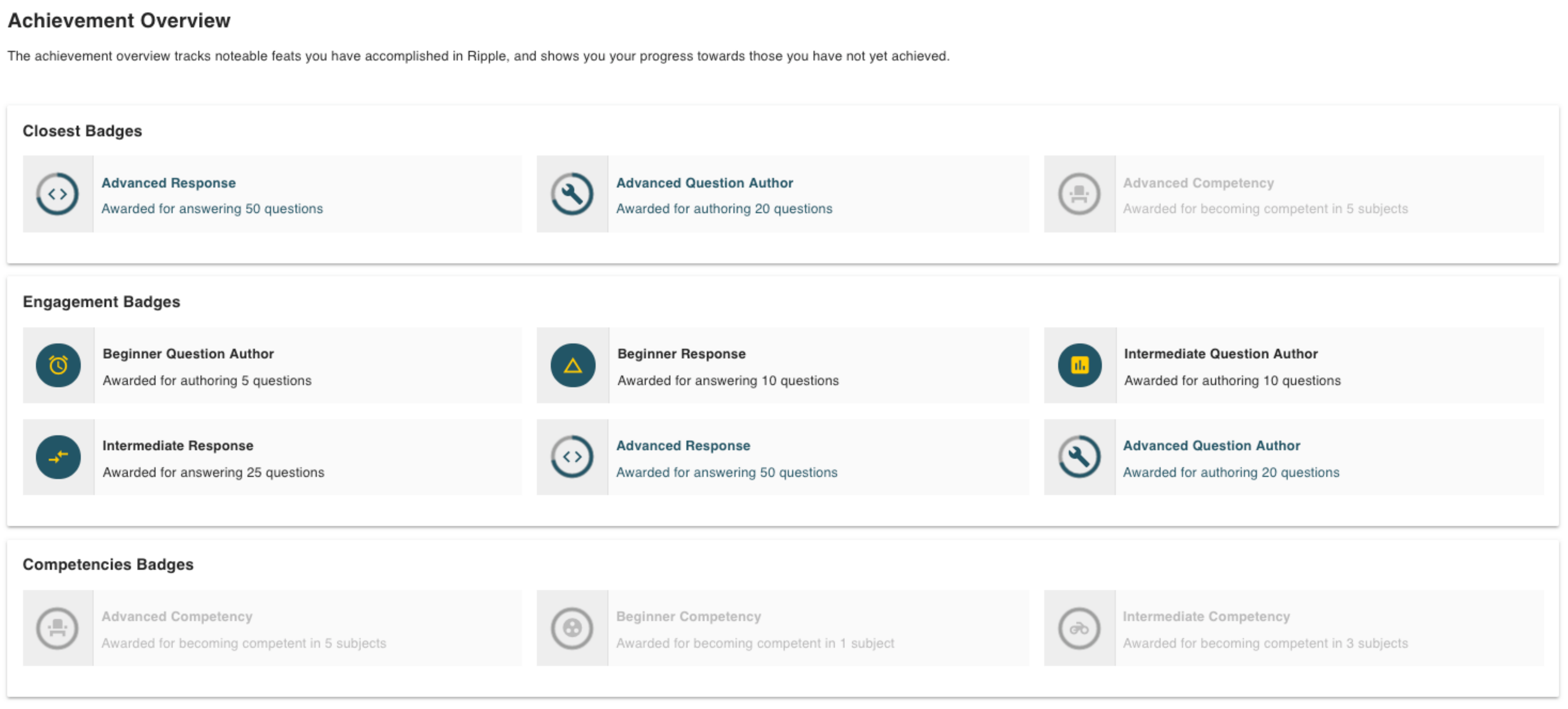}
\caption{The graphical interface used for showing achievements in \name. \label{fig:achievementVis}}
\end {figure*}

\paragraph{Achievements}
\name uses Gamification and badging to increase student motivation and performance. Students are able to achieve badges in three broad categories of ``Engagement Badges", ``Competency Badges"and ``Peer Support Badges". The achievement view enables students to track their progress towards achievements. Figure~\ref{fig:achievementVis} shows the graphical interface used for showing achievements in \name.

\paragraph{Notifications}
The notification view in \name allows students to view notifications about their achievements and up coming study sessions. 

\paragraph{Consent} Upon the first use of the platform, students are presented with a consent form seeking their permission to use their data to improve our understanding of the learning process and to evaluate the effectiveness of the recommended content. The consent view enables students to change their response at any time.

\subsection{Instructor Page}
Users with ``instructor" role have access to an additional page that has three main views: the course overview, consent form, and reported questions.

\paragraph{Course Overview} This view enables instructors to add the set of topics that are to be used for tagging the questions. This list can be updated throughout the semester. This view also allows instructors to track the progress of each of the students that are enrolled in the course. Data related to their progress can be downloaded as a CSV or an SQL dump, which may be useful for educational data mining researchers. 
\paragraph{Consent Form} This view enables instructors to develop the content of the consent form, which is to be filled by the students.

\paragraph{Reported Questions} This view enables instructors to view, edit, and delete questions that have been flagged as inappropriate. 

\section{Expected Benefits}\label{sec:benefits}
This section discusses the multiple beneficial outcomes that \name provides for students, instructors, and educational data mining researchers. 

\name enables students to:
\begin{itemize}
\item \textit{think about ways of evaluating understanding and learning.} Designing questions requires students to think carefully about the topics of the course and focuses attention on the learning outcomes; choosing distractors requires students to consider misconceptions, ambiguity and possible interpretations of concepts; and writing explanations require students to express their understanding of a topic with as much clarity as possible, helping them develop their written communication skills and deepen their understanding \cite{Denny2008}.

\item \textit{identify their knowledge gaps.} Students often lack the requisite skills for making good decisions about what and how to study \cite{Biggs1999}, which can leave them undirected and time wasted. RiPPLE uses knowledge tracing algorithms to approximate knowledge states of students, enabling them to identify their knowledge gaps.

\item \textit{receive a more tailored learning experience.} Having course content that serves the needs of diverse student populations (e.g., those with differing academic ability, backgrounds, and generational expectations) is extremely challenging. RiPPLE allows students to receive a more personalised learning experience by recommending content based on their knowledge states.

\item \textit{compare themselves with their peers or against their target goals.} Students are often curious to know how they are performing compared to their peers or are keen to set personal goals and track their progress. RiPPLE uses knowledge tracing algorithms and visualisations that empower students to track their progress.

\item \textit{become more effective communicators.} RiPPLE promotes collaborative learning in a social, student-centred learning environment, where students learn to articulate their opinions, provide support for their views, and listen and relate to the views of others. Such learning communities lead to the development of ``cognitive or intellectual skills or to an increase in knowledge and understanding", or to the development of communication and professional skills \cite{falchikov2001}.

\item  \textit{enhance their social connectedness.} Many students, especially in their first year of university, will have difficulty navigating their new academic and social environment and adequately exploiting the available resources. RiPPLE enables students to grow their social connectedness, connecting students with peers who can adequately help each other for collaborative assessments or in reaching their academic aspirations. 

\item \textit{increase their digital literacy.} Through exposure to innovative visualisation approaches, students will develop an appreciation for methods of communication and externalisation of knowledge from complex data sets.
\end{itemize}

RiPPLE enables instructors to:
\begin{itemize}
\item \textit{utilize crowdsourcing to develop course content. }Implementing adaptively in a course requires a large amount of new content and learning object tagging. Through the use of crowdsourcing, RiPPLE enables instructors to develop a large data set of tagged course content.
\item \textit{provide rich and immediate feedback to students.} Instructors find it challenging to provide meaningful, rich and timely feedback at scale. RiPPLE provides immediate feedback to students on their progress and provides recommendations to help them overcome their knowledge gaps.

\item \textit{identify individual-level and course-level knowledge gaps. }Instructors often find it challenging, especially in large classes, to comprehend individual-level and course-level gaps. RiPPLE informs instructors of these gaps, so that they can update their course content accordingly.

\item \textit{identify at-risk students early in the semester.} Instructors find it challenging to identify at-risk students in large classes, while working within limited budgets. RiPPLE uses knowledge tracing algorithms to identify at-risk students.
\end{itemize}

Finally, RiPPLE enables learning analytics and educational data mining researchers to develop and validate their own knowledge tracing and recommender system algorithms. Currently, most researchers have to validate these algorithms using synthetic or historical data sets that do not provide compelling evidence that they lead to better learning. RiPPLE enables researchers to validate their algorithms in a live setting using parallel-group double-blind randomised trials or A/B Testing to determine whether recommendations lead to measurable gains.

\section{Implementation Details of the RiPPLE Client}\label{sec:client}
This section provides information on the implementation of the RiPPLE Client. Further details can be found on the Wiki\footnote{\url{https://github.com/hkhosrav/RiPPLE-Core/wiki}} page of the project on GitHub.

The RiPPLE client is a VueJS \footnote{\url{https://vuejs.org/}} application augmented with Awesome Vue TS\footnote{\url{https://github.com/HerringtonDarkholme/av-ts}} for the Typescript language. Its primary purpose is to consume the RiPPLE API supported by the RiPPLE Server (presented in Section~\ref{sec:server}) to provide its information to the user.

\subsection{Source Code Conventions and Overview of the \name Client}
The RiPPLE client follows several conventions to ensure the scalability and managability of the software. Notably, it follows the Service and Repository, and Subscription patterns to ensure development consistency and reduce code duplication. It includes the following folders:
\begin{itemize}
\item The \textit{project root} only contains files which are used for starting the build process, core application configuration, and files to dictate the behaviour of external tools interacting with the project (such as git, typescript, and npm).
\item The \textit{build directory} contains files required for the build process. This is a strange convention, but has been made popular by the Vue community and adopted by many JavaScript developers.
\item The \textit{config directory} contains files which are used to influence the build process. 
\item The\textit{ dist directory} is short for distributable. It contains the build files required for a production build. Files created from a development build are not placed here, and are instead stored in memory.
\item The \textit{docs directory} contains the same thing as the dist directory for each merge into master. This is to ensure that GitHub will serve the latest client build.

\item The \textit{node\_modules directory} contains 3rd party project dependencies created by npm. It should not be in source control

\item The \textit{src directory} contains source files for the application.
\item The \textit{test directory} contains the specification verification of how the application operates. They are executed with \textit{npm run test}.
\end{itemize}

Within the src directory, there are a few subdirectories. The \textit{components directory} contains all of the views for the application. In this case, VueJS is acting as the view. The components directory is subdivided into multiple directories based on the purpose of the component. The \textit{interfaces directory} contains the TypeScript interfaces for the application. These are used heavily throughout the application and typically mirror the \textit{toJSON()} method of the RiPPLE Server models. The \textit{repositories directory} contains entities which are responsible for retrieving the external data required by the application. It is encapsulated into a single place because it (1) provides a substitution point for testing (2) allows for a flexible system architecture for easier 'plugging-in' to changing parts (e.g., mocking data on client vs. retrieving from server), and (3) is consistent and predictable, which in turn provides for a better developer experience. Typically, the repositories are only interacted with from the services. This top-down approach is consistent with the Vue framework. the \textit{routes directory} defines the routes of an application such that a URL will map to a Vue component. This is necessary in single-page applications which are dependent on state. The \textit{services directory} contains a service layer which aims to encapsulate all of the business logic of the application. Given the asynchronous nature of the application, a subscription system called Fetcher also exists here. Finally, the \textit{style directory} contains global CSS styles for the application. 

\subsection{Deployment of the RiPPLE Client}
The RiPPLE client architecture is relatively simple, however it does need some configuration to work correctly.

\subsubsection{Installing Dependencies} From your terminal, make sure you can access both nodeJS (\textit{node --version}) and npm (\textit{npm --version}). If not, you will be unable to continue. This project ships with a \textit{package.json} file, which details all of the modules required to build and run the application. The simplest way to install them all is to run \textit{npm install} from your terminal. 

npm will automatically read the package.json file and install everything it requires. Generally you only need to do this once, but if new dependencies are added you will need to re-run \textit{npm install}.

\subsubsection{General Configuration} RiPPLE client side will read its configuration from what is called an \textit{environment variable}. These variables are injected into the application at runtime by replacing all instances of \textit{process.env.*} with the name of predefined variable. These predefined variables are things such as debug mode and where to look for the API. Since environment variables can sometimes be sensitive (such as database passwords), it is best practice to keep them out of source control. 

RiPPLE reads its environment variables from environment files, which are named \textit{.env.dev}, and \textit{.env.prod} for development and production respectively. In a fresh install, the .env.dev and .env.prod files will not exist (since they are not in source control). Instead, a .env.example file is provided which has all of the possible environment variable names, and example values for them. 

If you make your own .env.dev or .env.prod files, you can place whatever you like in it, and your git client will automatically ignore anything inside of it. Supported environment values include the following:

\begin{itemize}

\item API\_LOCATION - string: A URI which points to the RiPPLE Server. Should not have a trailing '/'.
\item Node\_ENV - string: A string indicating the development environment as PRODUCTION or DEVELOPMENT.
\end{itemize}

\subsubsection{Development Configuration} Typically, a RiPPLE client in development mode will connect to a Development RiPPLE Server (See Section~\ref{sec:server}). By default, this will be http://localhost:9000. You will need to initialize your .env.dev environment file to development values. Once your environment file has been created and your [dependencies installed ]RiPPLE-Client---Environment) you will be able to build and test your application. The key command for active development is \textit{npm run dev}, which builds your application and opens it in your browser.

\paragraph{Development Commands} This project comes packaged with several convenience commands
\begin{itemize}
\item \textit{npm run dev}: Creates a local development server, and serves your files over it. When a local file is changed your project will be rebuilt incrementally and the new code injected into your web browser.
\item \textit{npm run lint}: Runs eslint over the codebase to ensure it meets the specified JavaScript style guide.
\item \textit{npm run unit}: Runs all unit tests for the project. They are located under ./test/unit/spec/
\item \textit{npm run e2e}: Runs all integration tests for the project. They are located under ./test/e2e/spec/. The current webdriver used is PhantomJS (runs headlessly). It is installed locally as part of \textit{npm install}.
\item \textit{npm run test}: Runs all unit tests and e2e tests. It is the same as running both \textit{npm run unit} and  \textit{npm run e2e}.
\end{itemize}

\paragraph{Helpful Development Tools} The Chrome Console\footnote{\url{https://developers.google.com/web/tools/chrome-devtools/console/}} allows you to inspect your application at runtime. All errors will also be reported to this console which allows for easy issue identification. The console also has a debugger, which is very useful for inspecting code to make sure it behaves correctly. The Vue Devtools\footnote{\url{https://github.com/vuejs/vue-devtools}} allows inspection of Vue components.

\subsubsection{Production Configuration} Typically, a RiPPLE client in production mode will connect to a Production RiPPLE Server, which needs to be manually configured. You will need to initialize your .env.prod environment file to production values. An example environment file is provided in the repository, but the defaults are not suitable for a production application. Once your environment file has been created and your dependencies installed you will be able to build your application. The key commands for building is \textit{npm run test} and then \textit{npm run build}, which builds your application and places it into the \textit{dist directory}. This directory is on the .gitignore, so any changes to it will not be committed to the project.

Once you have built your application, you will need to serve it via an actual webserver. This is most commonly done by placing it into /var/www/htdocs/ of a machine configured to run nginx\footnote{\url{https://nginx.org/en/}}. 

This GitHub project is configured to serve the \textit{docs directory} of the master branch on demo page\footnote{\url{https://hkhosrav.github.io/RiPPLE-Core/\#/}}. You should therefore remember to copy the result of ./dist into ./docs before a merge into master to ensure that the latest version is being served via GitHub.

\section{Implementation Details of the RiPPLE Server} \label{sec:server} \label{sec:server}

This section provides information on the implementation of the RiPPLE Sever. Further details can be found on the Wiki\footnote{\url{https://github.com/hkhosrav/RiPPLE-Core/wiki}} page of the project on GitHub.

The RiPPLE server is a Django\footnote{\url{https://www.djangoproject.com/}} application running on Python. Its primary purpose is to provide a data gateway to user information, authentication through LTI, and personalised recommendations for the student competency

\subsection{Source Code Conventions and Overview of the \name Server}
The RiPPLE server follows several conventions to ensure the scalability and managability of the software. Notably, it follows the Service and Repository, as well as Django-community favoured patterns to ensure development consistency and reduce code duplication. It includes the following folders:

\begin{itemize}
\item The\textit{ project root} only contains project-specific files which are used for getting started in the project. This includes things like dependencies, licensing, and a production deployment scripts
\item The \textit{src directory} contains source files for the application. You will spend lots of time in this directory. It follows the Django Project Structure, and should contain environment files and Django bootstrap files (such as manage.py).
\item The \textit{ripple directory} contains the core application configuration. This directory contains files important configuration files such as settings.py, urls.py which are necessary to tie different apps together.

\item The \textit{src directory} contains many subfolders, each of which is refereed to as a Django App\footnote{\url{https://docs.djangoproject.com/en/1.11/intro/tutorial01/\#creating-the-polls-app}}. To quote from Django: "What’s the difference between a project and an app? An app is a Web application that does something – e.g., a Weblog system, a database of public records or a simple poll app. A project is a collection of configuration and apps for a particular website. A project can contain multiple apps. An app can be in multiple projects." In order to promote code re-usability; the RiPPLE server makes use of projects where appropriate.

\item The \textit{services directory} inside each app contains entities which are responsible for handling the business logic of the application. They should be used exclusively by the routing controllers to ensure that tasks are always completed in the same way (e.g., using the QuestionSearch is beneficial since it will automatically ensure your search results are within the current course context). Services are encapsulated into a single place. 
\end{itemize}

Each App has the following files
\begin{itemize}
\item \textit{urls.py}, which contains application route definitions. Routes are not automatically added to the global application, you must also modify \textit{ripple/urls.py} to import your app routes.
\item \textit{views.py}, which contains the application controllers.
\item tests.py, which contains the specification verification of how the application operates. They are executed with \textit{python manage.py test}.
\end{itemize}

\subsection{Deployment of the RiPPLE Server}
The RiPPLE server architecture is relatively simple since it builds off of one of the most popular python frameworks, however it does need some configuration to work correctly.

\subsubsection{Installing Dependencies}
From your terminal, make sure you can access both python (python--version) and pip (pip --version). If not, you will be unable to continue.

It is highly recommended to run python through virtualenv to ensure you are developing in an isolated environment. Django recommends doing this

This project ships with a \textit{requirements.txt} file, which details all of the modules required to build and run the application. The simplest way to install them all is to run \textit{pip install -r requirements.txt} from your terminal at the project root. pip will automatically read the requirements.txt file and install everything it requires. Generally you only need to do this once, but if new dependencies are added you will need to re-run \textit{pip install -r requirements.txt}.

\subsubsection{General Configuration}

The RiPPLE server will read its configuration from what is called an \textit{environment variable}. These variables are injected into the application at runtime by replacing reading from \textit{os.getenv()}. These predefined variables are things such as debug mode and application secrets. Since environment variables can sometimes be sensitive (such as database passwords), it is best practice to keep them out of source control.

RiPPLE reads its environment variables from environment files, which are named \textit{.env.dev}, and \textit{.env.prod} for development and production respectively. 

In a fresh install, the \textit{.env.dev}, and \textit{.env.prod} files will not exist (since they are not in source control). Instead, an \textit{.env.example} file is provided which has all of the possible environment variable names, and example values for them. If you make your own \textit{.env.dev}, or\textit{ .env.prod} files, you can place whatever you like in it, and your git client will automatically ignore anything inside of it. Supported environment values include the following:

\begin{itemize}

\item API\_LOCATION - string: A URI which points to the RiPPLE Server. Should not have a trailing '/'.

\item DEVELOPMENT\_ENVIRONMENT - string: A string indicating the development environment, e.g., (PRODUCTION or DEVELOPMENT)

\item DJANGO\_KEY - string: The SECRET\_KEY environment variable required by django. It should be a large unique string

\item PROXY\_LOCATION - string: The subpath of the proxy\_pass location if the application is being run through nginx and is not on the root path.

\item LTI\_SUCCESS\_REDIRECT - string: URI to redirect to after a successful LTI validation request
\item LTI\_URL - string: URI to use in LTI validation
\item LTI\_APP\_KEY - string: Application key to pass to LTI validation service
\item DATABASE\_TYPE - string: Database type to use (e.g., mysql, sqlite3)
\item DATABASE\_NAME - string: Name of database to use
\item DATABASE\_HOST - string: Host of database
\item DATABASE\_USER - string: Username to use when connecting to database
\item DATABASE\_PASSWORD - string: Password to use authenticate with on database
\end{itemize}

\subsubsection{Development Configuration}
Typically, a RiPPLE server in development mode will be running on the loopback address of the machine; this is http://localhost:9000 by default. 

To initialise your development environment, you have two options: (1) run \textit{python manage.py env dev} - which will prompt to create a .env.dev if none exists or (2) run \textit{cp .env.example .env.dev} - which will copy the example configuration into a development file. 

Once your environment file has been created and your dependencies installed you will be able to build and test your application. The key command for active development is \textit{python manage.py runserver}, which builds your application and enables hot-reload.

From a fresh install, the development workflow is typically:
\begin{enumerate}
\item \textit{pip install -r requirements.txt} - Install project dependencies
\item \textit{cd src} - Change to project working directory
\item \textit{CREATE\_SCHEMA} - If not using sqlite3, then the schema must manually be created
\item \textit{python manage.py migrate} - Creates database schema on machine
\item \textit{python manage.py seed} - Populates database with values
\item \textit{python manage.py seedCourse --name courseName --course corseCode --file /path/to/JSONfile --host /host/domain -} Populates database with values from file
\item \textit{python manage.py runserver} - Starts the server
\item edit files and make changes
\end{enumerate}

\paragraph{Running in unauthenticated mode} If you wish to run the application without authentication, change the ALLOW\_UNAUTHENTICATED parameter in src/ripple/settings.py to True. This will disable token verification on login requests, and instead assign the requester a random user token when the /users/login endpoint is accessed. please note that the unauthenticated mode should never be enabled on a live system with real users.

\paragraph{Development Commands}
\begin{itemize}
\item \textit{python manage.py test}: Runs the application unit tests
\item \textit{python manage.py makemigrations}: Reads in all changes from project models and creates migration files from them
\item \textit{python manage.py migrate}: Updates the database schema to match the migration definitions
\item \textit{python manage.py seed} Seeds the database with mock data
\item \textit{python manage.py env ENV\_NAME} ENV\_NAME must be either "prod" or "dev".It reads in the environment definition from \textit{.env.prod} or \textit{.env.dev} respectively, and makes it the active environment
\item \textit{python manage.py runserver}: Spawns a webserver on http://localhost:9000 running the application with hot-reload
\item \textit{python manage.py runsslserver}: Spawns a HTTPS compatible server on https://localhost:9000 running the application with hot-reload
\end{itemize}

\subsubsection{Production Configuration}
Typically, a RiPPLE server in development mode will be running on the loopback address of the machine through the use of a stable webserver (e.g., gunicorn), but with an external web server (e.g., nginx) proxying the request into the loopback adrress.

You will need to initialize your \textit{.env.prod} environment file to production values, which should be kept secret. An example .env file is provided in the repository (.env.example), but\textit{ python manage.py env ENV\_NAME} will also create an \textit{.env.current} file if none exists.

Ensure that \textit{./src/ripple/settings.py} has ALLOW\_UNAUTHENTICATED set to False if you are deploying to a live system

To initialise your production environment, you have two options: (1) run \textit{python manage.py prod} - which will prompt to create a \textit{.production.env} if none exists amd (2) run \textit{cp .env.example .env.prod} - which will copy the example configuration into a development file

Once your environment file has been created and your dependencies installed you will be able to build and test your application. The key command for production development is \textit{sudo -E ./deploy.sh .env.prod}, which will deploy your application and run it as a service. Running the application as a service is the best way to ensure it continues to run after an SSH disconnect

\section{Contributing}\label{sec:contributing}
\name is an open-source system - and always will be. Collaboration, improvements, suggestions and pull requests are always welcome. In order to streamline the development process; the following guide exists to help people get started:

\begin{enumerate}
\item Open a ticket in the ticketing system - this will serve as a point of contact to the project maintainers.
\item Either: a. Create a branch of the format RIPPLE-\#:ticketId. (eg. RIPPLE-\#33 for ticket \#33). b. Fork the project and do your work there.
\item When the work is finished, create a pull request to indicate your changes are ready to be reviewed. In your pull request; document your changes and preferably link to your ticket.
\item Your code will be reviewed by a project contributor, and will be merged in if it aligns with the projects goals (which was hopefully discussed previously in your ticket).
\item After your code has been merged in, this documentation will be updated as appropriate.
\end{enumerate}

\bibliographystyle{acmtrans}
\bibliography{Master-references} 	

\begin{thebibliography}{}

\bibitem[\protect\citeauthoryear{Baker, Corbett, and Aleven}{Baker
  et~al\mbox{.}}{2008}]{Baker2008}
{\sc Baker, R.}, {\sc Corbett, A.~T.}, {\sc and} {\sc Aleven, V.} 2008.
\newblock More accurate student modeling through contextual estimation of slip
  and guess probabilities in bayesian knowledge tracing.
\newblock In {\em International Conference on Intelligent Tutoring Systems}.
  Springer, 406--415.

\bibitem[\protect\citeauthoryear{Banfield and Wilkerson}{Banfield and
  Wilkerson}{2014}]{banfield2014increasing}
{\sc Banfield, J.} {\sc and} {\sc Wilkerson, B.} 2014.
\newblock Increasing student intrinsic motivation and self-efficacy through
  gamification pedagogy.
\newblock {\em Contemporary Issues in Education Research (Online)\/}~{\em
  7,\/}~4, 291.

\bibitem[\protect\citeauthoryear{Biggs}{Biggs}{1999}]{Biggs1999}
{\sc Biggs, J.} 1999.
\newblock What the student does: teaching for enhanced learning.
\newblock {\em Higher education research \& development\/}~{\em 18,\/}~1,
  57--75.

\bibitem[\protect\citeauthoryear{Bovill, Cook-Sather, and Felten}{Bovill
  et~al\mbox{.}}{2011}]{bovill2011students}
{\sc Bovill, C.}, {\sc Cook-Sather, A.}, {\sc and} {\sc Felten, P.} 2011.
\newblock Students as co-creators of teaching approaches, course design, and
  curricula: implications for academic developers.
\newblock {\em International Journal for Academic Development\/}~{\em 16,\/}~2,
  133--145.

\bibitem[\protect\citeauthoryear{Cai, Bain, Krzywicki, Wobcke, Kim, Compton,
  and Mahidadia}{Cai et~al\mbox{.}}{2011}]{cai2011learning}
{\sc Cai, X.}, {\sc Bain, M.}, {\sc Krzywicki, A.}, {\sc Wobcke, W.}, {\sc Kim,
  Y.~S.}, {\sc Compton, P.}, {\sc and} {\sc Mahidadia, A.} 2011.
\newblock Learning to make social recommendations: a model-based approach.
\newblock In {\em International Conference on Advanced Data Mining and
  Applications}. Springer, 124--137.

\bibitem[\protect\citeauthoryear{Cazella, Reategui, and Behar}{Cazella
  et~al\mbox{.}}{2010}]{Cazella2010}
{\sc Cazella, S.}, {\sc Reategui, E.}, {\sc and} {\sc Behar, P.} 2010.
\newblock Recommendation of learning objects applying collaborative filtering
  and competencies.
\newblock In {\em Key Competencies in the Knowledge Society}. Springer, 35--43.

\bibitem[\protect\citeauthoryear{Cechinel, Sicilia, S{\'a}Nchez-Alonso, and
  Garc{\'\i}A-Barriocanal}{Cechinel et~al\mbox{.}}{2013}]{Cechinel2013}
{\sc Cechinel, C.}, {\sc Sicilia, M.-{\'A}.}, {\sc S{\'a}Nchez-Alonso, S.},
  {\sc and} {\sc Garc{\'\i}A-Barriocanal, E.} 2013.
\newblock Evaluating collaborative filtering recommendations inside large
  learning object repositories.
\newblock {\em Information Processing \& Management\/}~{\em 49,\/}~1, 34--50.

\bibitem[\protect\citeauthoryear{Cen, Koedinger, and Junker}{Cen
  et~al\mbox{.}}{2006}]{cen2006learning}
{\sc Cen, H.}, {\sc Koedinger, K.}, {\sc and} {\sc Junker, B.} 2006.
\newblock Learning factors analysis-a general method for cognitive model
  evaluation and improvement.
\newblock In {\em Intelligent tutoring systems}. Vol. 4053. Springer, 164--175.

\bibitem[\protect\citeauthoryear{Chambers, Cleveland, Kleiner, Tukey,
  et~al\mbox{.}}{Chambers et~al\mbox{.}}{1983}]{chambers1983graphical}
{\sc Chambers, J.~M.}, {\sc Cleveland, W.~S.}, {\sc Kleiner, B.}, {\sc Tukey,
  P.~A.}, {\sc et~al\mbox{.}} 1983.
\newblock {\em Graphical methods for data analysis}. Vol.~5.
\newblock Wadsworth Belmont, CA.

\bibitem[\protect\citeauthoryear{Chen}{Chen}{2008}]{chen2008}
{\sc Chen, C.-M.} 2008.
\newblock Intelligent web-based learning system with personalized learning path
  guidance.
\newblock {\em Computers \& Education\/}~{\em 51,\/}~2, 787--814.

\bibitem[\protect\citeauthoryear{Chen and Nayak}{Chen and
  Nayak}{2013}]{chen2013}
{\sc Chen, L.} {\sc and} {\sc Nayak, R.} 2013.
\newblock A reciprocal collaborative method using relevance feedback and
  feature importance.
\newblock In {\em Proceedings of the 2013 IEEE/WIC/ACM International Joint
  Conferences on Web Intelligence (WI) and Intelligent Agent Technologies (IAT)
  - Volume 01}. WI-IAT '13. IEEE Computer Society, Washington, DC, USA,
  133--138.

\bibitem[\protect\citeauthoryear{Cooper and Khosravi}{Cooper and
  Khosravi}{2018}]{Cooper2018}
{\sc Cooper, K.} {\sc and} {\sc Khosravi, H.} 2018.
\newblock Graph-based visual topic dependency models.
\newblock In {\em Proceedings of the 8th International Conference on Learning
  Analytics and Knowledge}.

\bibitem[\protect\citeauthoryear{Corbett}{Corbett}{2001}]{corbett2001cognitive}
{\sc Corbett, A.} 2001.
\newblock Cognitive computer tutors: Solving the two-sigma problem.
\newblock {\em User Modeling 2001\/}, 137--147.

\bibitem[\protect\citeauthoryear{Corbett and Anderson}{Corbett and
  Anderson}{1994}]{corbett1994knowledge}
{\sc Corbett, A.~T.} {\sc and} {\sc Anderson, J.~R.} 1994.
\newblock Knowledge tracing: Modeling the acquisition of procedural knowledge.
\newblock {\em User modeling and user-adapted interaction\/}~{\em 4,\/}~4,
  253--278.

\bibitem[\protect\citeauthoryear{Denny, Hamer, Luxton-Reilly, and
  Purchase}{Denny et~al\mbox{.}}{2008}]{Denny2008}
{\sc Denny, P.}, {\sc Hamer, J.}, {\sc Luxton-Reilly, A.}, {\sc and} {\sc
  Purchase, H.} 2008.
\newblock Peerwise: students sharing their multiple choice questions.
\newblock In {\em Proceedings of the fourth international workshop on computing
  education research}. ACM, 51--58.

\bibitem[\protect\citeauthoryear{Drachsler, Verbert, Santos, and
  Manouselis}{Drachsler et~al\mbox{.}}{2015}]{Drachsler2015}
{\sc Drachsler, H.}, {\sc Verbert, K.}, {\sc Santos, O.~C.}, {\sc and} {\sc
  Manouselis, N.} 2015.
\newblock Panorama of recommender systems to support learning.
\newblock In {\em Recommender systems handbook}. Springer, 421--451.

\bibitem[\protect\citeauthoryear{Erdt, Fern{\'a}ndez, and Rensing}{Erdt
  et~al\mbox{.}}{2015}]{Erdt2015}
{\sc Erdt, M.}, {\sc Fern{\'a}ndez, A.}, {\sc and} {\sc Rensing, C.} 2015.
\newblock Evaluating recommender systems for technology enhanced learning: A
  quantitative survey.
\newblock {\em IEEE Transactions on Learning Technologies\/}~{\em 8,\/}~4,
  326--344.

\bibitem[\protect\citeauthoryear{Essa}{Essa}{2016}]{essa2016possible}
{\sc Essa, A.} 2016.
\newblock A possible future for next generation adaptive learning systems.
\newblock {\em Smart Learning Environments\/}~{\em 3,\/}~1, 16.

\bibitem[\protect\citeauthoryear{Falchikov}{Falchikov}{2001}]{falchikov2001}
{\sc Falchikov, N.} 2001.
\newblock {\em Learning together: Peer tutoring in higher education}.
\newblock Psychology Press.

\bibitem[\protect\citeauthoryear{Falmagne, Cosyn, Doignon, and
  Thi{\'e}ry}{Falmagne et~al\mbox{.}}{2006}]{falmagne2006assessment}
{\sc Falmagne, J.-C.}, {\sc Cosyn, E.}, {\sc Doignon, J.-P.}, {\sc and} {\sc
  Thi{\'e}ry, N.} 2006.
\newblock The assessment of knowledge, in theory and in practice.
\newblock In {\em Formal concept analysis}. Springer, 61--79.

\bibitem[\protect\citeauthoryear{Fazeli, Loni, Drachsler, and Sloep}{Fazeli
  et~al\mbox{.}}{2014}]{Fazeli2014}
{\sc Fazeli, S.}, {\sc Loni, B.}, {\sc Drachsler, H.}, {\sc and} {\sc Sloep,
  P.} 2014.
\newblock Which recommender system can best fit social learning platforms?
\newblock In {\em European Conference on Technology Enhanced Learning}.
  Springer, 84--97.

\bibitem[\protect\citeauthoryear{Gomez-Albarran and
  Jimenez-Diaz}{Gomez-Albarran and Jimenez-Diaz}{2009}]{Gomez-Albarran2009}
{\sc Gomez-Albarran, M.} {\sc and} {\sc Jimenez-Diaz, G.} 2009.
\newblock Recommendation and students’ authoring in repositories of learning
  objects: A case-based reasoning approach.
\newblock {\em International Journal of Emerging Technologies in Learning
  (iJET)\/}~{\em 4,\/}~2009, 35--40.

\bibitem[\protect\citeauthoryear{Guy}{Guy}{2015}]{Guy2015}
{\sc Guy, I.} 2015.
\newblock Social recommender systems.
\newblock In {\em Recommender Systems Handbook}. Springer, 511--543.

\bibitem[\protect\citeauthoryear{Hong, Zheng, Wang, and Shi}{Hong
  et~al\mbox{.}}{2013}]{Hong2013}
{\sc Hong, W.}, {\sc Zheng, S.}, {\sc Wang, H.}, {\sc and} {\sc Shi, J.} 2013.
\newblock A job recommender system based on user clustering.
\newblock {\em Journal of Computers\/}~{\em 8,\/}~8, 1960--1967.

\bibitem[\protect\citeauthoryear{Imran, Belghis-Zadeh, Chang, Graf,
  et~al\mbox{.}}{Imran et~al\mbox{.}}{2016}]{Imran2016}
{\sc Imran, H.}, {\sc Belghis-Zadeh, M.}, {\sc Chang, T.-W.}, {\sc Graf, S.},
  {\sc et~al\mbox{.}} 2016.
\newblock Plors: a personalized learning object recommender system.
\newblock {\em Vietnam Journal of Computer Science\/}~{\em 3,\/}~1, 3--13.

\bibitem[\protect\citeauthoryear{Jose}{Jose}{2016}]{Ferreira2016white}
{\sc Jose, F.} 2016.
\newblock White paper: Knewton adaptive learning building the world’s most
  powerful recommendation engine for education.

\bibitem[\protect\citeauthoryear{Khajah, Huang, Gonz{\'a}lez-Brenes, Mozer, and
  Brusilovsky}{Khajah et~al\mbox{.}}{2014}]{khajah2014integrating}
{\sc Khajah, M.~M.}, {\sc Huang, Y.}, {\sc Gonz{\'a}lez-Brenes, J.~P.}, {\sc
  Mozer, M.~C.}, {\sc and} {\sc Brusilovsky, P.} 2014.
\newblock Integrating knowledge tracing and item response theory: A tale of two
  frameworks.
\newblock In {\em Proceedings of Workshop on Personalization Approaches in
  Learning Environments (PALE 2014) at the 22th International Conference on
  User Modeling, Adaptation, and Personalization}. University of Pittsburgh,
  7--12.

\bibitem[\protect\citeauthoryear{Khosravi, Cooper, and Kitto}{Khosravi
  et~al\mbox{.}}{2017}]{khosravi2017}
{\sc Khosravi, H.}, {\sc Cooper, K.}, {\sc and} {\sc Kitto, K.} 2017.
\newblock Riple: Recommendation in peer-learning environments based on
  knowledge gaps and interests.
\newblock {\em JEDM-Journal of Educational Data Mining\/}~{\em 9,\/}~1, 42--67.

\bibitem[\protect\citeauthoryear{Kopeinik, Lex, Seitlinger, Albert, and
  Ley}{Kopeinik et~al\mbox{.}}{2017}]{Kopeinik2017}
{\sc Kopeinik, S.}, {\sc Lex, E.}, {\sc Seitlinger, P.}, {\sc Albert, D.}, {\sc
  and} {\sc Ley, T.} 2017.
\newblock Supporting collaborative learning with tag recommendations: a
  real-world study in an inquiry-based classroom project.
\newblock In {\em Proceedings of the Seventh International Learning Analytics
  \& Knowledge Conference}. ACM, 409--418.

\bibitem[\protect\citeauthoryear{Kutty, Nayak, and Chen}{Kutty
  et~al\mbox{.}}{2014}]{kutty2014}
{\sc Kutty, S.}, {\sc Nayak, R.}, {\sc and} {\sc Chen, L.} 2014.
\newblock A people-to-people matching system using graph mining techniques.
\newblock {\em World Wide Web\/}~{\em 17,\/}~3, 311--349.

\bibitem[\protect\citeauthoryear{Landers and Landers}{Landers and
  Landers}{2014}]{landers2014empirical}
{\sc Landers, R.~N.} {\sc and} {\sc Landers, A.~K.} 2014.
\newblock An empirical test of the theory of gamified learning: The effect of
  leaderboards on time-on-task and academic performance.
\newblock {\em Simulation \& Gaming\/}~{\em 45,\/}~6, 769--785.

\bibitem[\protect\citeauthoryear{Lemire, Boley, McGrath, and Ball}{Lemire
  et~al\mbox{.}}{2005}]{Lemire2005}
{\sc Lemire, D.}, {\sc Boley, H.}, {\sc McGrath, S.}, {\sc and} {\sc Ball, M.}
  2005.
\newblock Collaborative filtering and inference rules for context-aware
  learning object recommendation.
\newblock {\em Interactive Technology and Smart Education\/}~{\em 2,\/}~3,
  179--188.

\bibitem[\protect\citeauthoryear{Mangina and Kilbride}{Mangina and
  Kilbride}{2008}]{Mangina2008}
{\sc Mangina, E.} {\sc and} {\sc Kilbride, J.} 2008.
\newblock Evaluation of keyphrase extraction algorithm and tiling process for a
  document/resource recommender within e-learning environments.
\newblock {\em Computers \& Education\/}~{\em 50,\/}~3, 807--820.

\bibitem[\protect\citeauthoryear{May, George, and Pr{\'e}v{\^o}t}{May
  et~al\mbox{.}}{2011}]{may2011travis}
{\sc May, M.}, {\sc George, S.}, {\sc and} {\sc Pr{\'e}v{\^o}t, P.} 2011.
\newblock Travis to enhance online tutoring and learning activities: Real-time
  visualization of students tracking data.
\newblock {\em Interactive Technology and Smart Education\/}~{\em 8,\/}~1,
  52--69.

\bibitem[\protect\citeauthoryear{Oxman, Wong, and Innovations}{Oxman
  et~al\mbox{.}}{2014}]{oxman2014white}
{\sc Oxman, S.}, {\sc Wong, W.}, {\sc and} {\sc Innovations, D.} 2014.
\newblock White paper: Adaptive learning systems.
\newblock {\em Integrated Education Solutions\/}.

\bibitem[\protect\citeauthoryear{Paramythis and Loidl-Reisinger}{Paramythis and
  Loidl-Reisinger}{2003}]{paramythis2003adaptive}
{\sc Paramythis, A.} {\sc and} {\sc Loidl-Reisinger, S.} 2003.
\newblock Adaptive learning environments and e-learning standards.
\newblock In {\em Second european conference on e-learning}. Vol.~1. 369--379.

\bibitem[\protect\citeauthoryear{Pardos and Heffernan}{Pardos and
  Heffernan}{2011}]{pardos2011kt}
{\sc Pardos, Z.} {\sc and} {\sc Heffernan, N.} 2011.
\newblock Kt-idem: introducing item difficulty to the knowledge tracing model.
\newblock {\em User Modeling, Adaption and Personalization\/}, 243--254.

\bibitem[\protect\citeauthoryear{Pavlik~Jr, Cen, and Koedinger}{Pavlik~Jr
  et~al\mbox{.}}{2009}]{pavlik2009performance}
{\sc Pavlik~Jr, P.~I.}, {\sc Cen, H.}, {\sc and} {\sc Koedinger, K.~R.} 2009.
\newblock Performance factors analysis--a new alternative to knowledge tracing.
\newblock {\em Online Submission\/}.

\bibitem[\protect\citeauthoryear{Piech, Bassen, Huang, Ganguli, Sahami, Guibas,
  and Sohl-Dickstein}{Piech et~al\mbox{.}}{2015}]{Piech2015}
{\sc Piech, C.}, {\sc Bassen, J.}, {\sc Huang, J.}, {\sc Ganguli, S.}, {\sc
  Sahami, M.}, {\sc Guibas, L.~J.}, {\sc and} {\sc Sohl-Dickstein, J.} 2015.
\newblock Deep knowledge tracing.
\newblock In {\em Advances in Neural Information Processing Systems}. 505--513.

\bibitem[\protect\citeauthoryear{Pizzato, Rej, Akehurst, Koprinska, Yacef, and
  Kay}{Pizzato et~al\mbox{.}}{2013}]{pizzato2013}
{\sc Pizzato, L.}, {\sc Rej, T.}, {\sc Akehurst, J.}, {\sc Koprinska, I.}, {\sc
  Yacef, K.}, {\sc and} {\sc Kay, J.} 2013.
\newblock Recommending people to people: the nature of reciprocal recommenders
  with a case study in online dating.
\newblock {\em User Modeling and User-Adapted Interaction\/}~{\em 23,\/}~5,
  447--488.

\bibitem[\protect\citeauthoryear{Pizzato, Rej, Chung, Yacef, Koprinska, and
  Kay}{Pizzato et~al\mbox{.}}{2010}]{pizzato2010reciprocal}
{\sc Pizzato, L.}, {\sc Rej, T.}, {\sc Chung, T.}, {\sc Yacef, K.}, {\sc
  Koprinska, I.}, {\sc and} {\sc Kay, J.} 2010.
\newblock Reciprocal recommenders.
\newblock In {\em 8th Workshop on Intelligent Techniques for Web
  Personalization and Recommender Systems, UMAP}.

\bibitem[\protect\citeauthoryear{Potts, Khosravi, Reidsema, Bakharia,
  Belonogoff, and Fleming}{Potts et~al\mbox{.}}{2018}]{potts2018reciprocal}
{\sc Potts, B.}, {\sc Khosravi, H.}, {\sc Reidsema, C.}, {\sc Bakharia, A.},
  {\sc Belonogoff, M.}, {\sc and} {\sc Fleming, M.} 2018.
\newblock Reciprocal peer recommendation for learning purposes.
\newblock In {\em Proceedings of the 8th International Conference on Learning
  Analytics and Knowledge}.

\bibitem[\protect\citeauthoryear{Salehi}{Salehi}{2013}]{Salehi2013}
{\sc Salehi, M.} 2013.
\newblock Application of implicit and explicit attribute based collaborative
  filtering and {BIDE} for learning resource recommendation.
\newblock {\em Data \& Knowledge Engineering\/}~{\em 87}, 130--145.

\bibitem[\protect\citeauthoryear{Sha and Hong}{Sha and
  Hong}{2017}]{sha2017neural}
{\sc Sha, L.} {\sc and} {\sc Hong, P.} 2017.
\newblock Neural knowledge tracing.
\newblock In {\em International Conference on Brain Function Assessment in
  Learning}. Springer, 108--117.

\bibitem[\protect\citeauthoryear{Sparrow}{Sparrow}{2016}]{Sparrow2016}
{\sc Sparrow, S.} 2016.
\newblock Smart sparrow - adaptive elearning platform.

\bibitem[\protect\citeauthoryear{Thai-Nghe, Drumond, Horv{\'a}th,
  Krohn-Grimberghe, Nanopoulos, and Schmidt-Thieme}{Thai-Nghe
  et~al\mbox{.}}{2011}]{Thai-Nghe2011}
{\sc Thai-Nghe, N.}, {\sc Drumond, L.}, {\sc Horv{\'a}th, T.}, {\sc
  Krohn-Grimberghe, A.}, {\sc Nanopoulos, A.}, {\sc and} {\sc Schmidt-Thieme,
  L.} 2011.
\newblock Factorization techniques for predicting student performance.
\newblock {\em Educational Recommender Systems and Technologies: Practices and
  Challenges\/}, 129--153.

\bibitem[\protect\citeauthoryear{Verbert, Drachsler, Manouselis, Wolpers,
  Vuorikari, and Duval}{Verbert et~al\mbox{.}}{2011}]{Verbert2011}
{\sc Verbert, K.}, {\sc Drachsler, H.}, {\sc Manouselis, N.}, {\sc Wolpers,
  M.}, {\sc Vuorikari, R.}, {\sc and} {\sc Duval, E.} 2011.
\newblock Dataset-driven research for improving recommender systems for
  learning.
\newblock In {\em Proceedings of the 1st International Conference on Learning
  Analytics and Knowledge}. ACM, 44--53.

\bibitem[\protect\citeauthoryear{Yilmaz}{Yilmaz}{2017}]{yilmaz2017effects}
{\sc Yilmaz, B.} 2017.
\newblock Effects of adaptive learning technologies on math achievement: A
  quantitative study of aleks math software.
\newblock Ph.D. thesis, University of Missouri-Kansas City.

\bibitem[\protect\citeauthoryear{Yudelson, Koedinger, and Gordon}{Yudelson
  et~al\mbox{.}}{2013}]{yudelson2013individualized}
{\sc Yudelson, M.~V.}, {\sc Koedinger, K.~R.}, {\sc and} {\sc Gordon, G.~J.}
  2013.
\newblock Individualized bayesian knowledge tracing models.
\newblock In {\em International Conference on Artificial Intelligence in
  Education}. Springer, 171--180.

\end{thebibliography}
\end{document}